\documentclass[a4paper,12pt]{article}
\usepackage{amsfonts,slashed}
\usepackage{url}
\usepackage{latexsym}
\usepackage{amsfonts}
\usepackage{epsfig}
\usepackage{latexsym,amssymb}
  \usepackage{amsmath,amssymb,amsthm}
\usepackage{ifpdf}
\ifx\pdfoutput\undefined
   \pdffalse
   \usepackage{cite}
 \else
   \pdfoutput=1
   \pdftrue
  \usepackage[pdftex]{hyperref}
\fi

\numberwithin{equation}{section}
\def\be{\begin{equation}}
\def\ee{\end{equation}}
\def\ba{\begin{array}}
\def\ea{\end{array}}

\newcommand{\bea}{\begin{eqnarray}}
\newcommand{\eea}{\end{eqnarray}}

\newcommand{\ft}[2]{{\textstyle\frac{#1}{#2}}}

\def\Re{\mathop{\rm Re}\nolimits}
\def\Im{\mathop{\rm Im}\nolimits}

\def\rmi{{\rm i}}
\def\rmd{{\rm d}}
\def\rme{{\rm e}}

\newcommand{\bbox}{\lower.2ex\hbox{$\Box$}}
\newcommand{\SU}{\mathop{\rm SU}}

\newcommand{\U}{\mathop{\rm {}U}}
\newcommand{\USp}{\mathop{\rm {}USp}}

\newcommand{\poin}{\Box\!\!\!\cdot}
\begin{document}
\begin{titlepage}
    \thispagestyle{empty}
    \begin{flushright}
        \hfill{CERN-PH-TH/2014-058}\\
        \hfill{DFPD-14/TH/05}
    \end{flushright}

    \vspace{15pt}

    \begin{center}
        { \LARGE{\bf A Search for an  $\mathcal{N}=2$ Inflaton Potential}}

        \vspace{20pt}

        {Anna Ceresole$^a$, Gianguido Dall'Agata$^{b}$, Sergio Ferrara $^{c,d,e}$,\\ Mario Trigiante$^{f}$ and Antoine Van Proeyen$^{g}$ }

        \vspace{25pt}

        {\small
        {\it ${}^a$ INFN, Sezione di Torino \\ Via Pietro Giuria 1, 10125 Torino, Italy}

        \vspace{15pt}

        {\it ${}^b$ Dipartimento di Fisica ``Galileo Galilei'' $\&$ INFN, Sezione di Padova \\
        Universit\`{a} di Padova, Via Marzolo 8, 35131 Padova, Italy}

        \vspace{15pt}

        {\it ${}^c$ Department of Physics, CERN Theory Division\\
        CH 1211, Geneva 23, Switzerland}

        \vspace{15pt}

        {\it ${}^d$ Department of Physics and Astronomy \\
        University of California, Los Angeles, CA, USA}}

\vspace{15pt}

        {\it ${}^e$ INFN - LNF,
        Via Enrico Fermi 40, I-00044 Frascati, Italy}

    \vspace{15pt}

{\it ${}^f$
Department of Applied Science and Technology,
Politecnico di Torino,
Corso Duca degli Abruzzi, 24
10129 Torino, Italy}

    \vspace{15pt}

{\it ${}^g$
Instituut voor Theoretische Fysica, KU Leuven,
Celestijnenlaan 200D, B-3001 Leuven, Belgium}

        \vspace{20pt}

        {ABSTRACT}

    \end{center}

We consider $\mathcal{N}=2$ supergravity theories that have the same spectrum as the $R+R^2$ supergravity, as predicted from the off-shell counting of degrees of freedom. These theories describe standard $\mathcal{N}=2$ supergravity coupled to one or two long massive vector multiplets. The central charge is not gauged in these models and they have a Minkowski vacuum with  $\mathcal{N}=2$ unbroken supersymmetry. The gauge symmetry, being non-compact, is always broken. $\alpha$-deformed inflaton potentials are obtained, in the case of a single massive vector multiplet, with $\alpha=1/3$ and $2/3$. The $\alpha=1$ potential (i.e. the Starobinsky potential) is also obtained, but only at the prize of having a single massive vector and a residual unbroken gauge symmetry. The inflaton corresponds to one of the Cartan fields  of the non-compact quaternionic-K\"{a}hler cosets.

    \vspace{10pt}

\end{titlepage}
\addtocounter{page}{1}
\tableofcontents
\newpage
\baselineskip 6 mm

\section{Introduction}
One of the attractive features of the inflaton potential of the Starobinsky model is its dual relation to a pure $R+R^2$ gravitational theory, where a physical scalar, the scalaron, emerges from the higher derivative theory. To understand this phenomenon in the context of supergravity, an off-shell formulation is needed, and in fact, two versions of the $R+R^2$ $\mathcal{N}=1$ supergravity were constructed \cite{Cecotti:1987sa,Cecotti:1987qe}. In these theories the inflaton is embedded in a massive chiral multiplet (\emph{old minimal} formulation) or in a massive vector multiplet  (\emph{new minimal} formulation) \cite{Farakos:2013cqa,Ferrara:2013rsa}. Recently, and supported by the PLANCK mission results \cite{Ade:2013zuv,Ade:2013uln}, cosmological properties and diverse extensions of these approaches have been widely considered in the literature.
More interestingly, the first formulation requires an additional chiral multiplet (other than the scalaron) which plays the role of the goldstino multiplet, which triggers the inflationary de Sitter phase at early times in the expansion of the Universe.
The more recent results of BICEP2 \cite{Ade:2014xna} seem to be at odds with the PLANCK ones and to disfavor Starobinsky-like models pointing towards more general, supergravity inspired inflationary scenarios. The apparent  tension between the two results is subject to considerable attention in the present literature. We shall refrain from dealing with this (still open) issue. Our analysis, although originally intended as an  exploration of the $\mathcal{N}=2$ extension of the $R+R^2$ gravity, yields results which can be relevant also to more general future investigations of string/supergravity-inspired inflationary models.\par
The superfield description of the $R+R^2$ theory in
the context of $\mathcal{N}=1$ and $\mathcal{N}=2$
supergravities have been considered by Ketov and collaborators \cite{Ketov:2013dfa,Ketov:2014qoa,Ketov:2014roa}. However, in
the $\mathcal{N}=2$ case, the proposed standard supergravity dual fails to describe the scalaron
multiplet and the corresponding potential for the reasons outlined in this paper. In order to embed $R+R^2$ gravity in an $\mathcal{N}=2$ setting, we use the correspondence between off-shell counting in Einstein supergravity and massive states in the corresponding higher-derivative theory. $\mathcal{N}=2$  supersymmetry puts much stronger restrictions on the hypothetical dual standard theory since massive multiplets can be either long or short BPS and one must give a criterion for the correct choice.  The off-shell formulation of $\mathcal{N}=2$ Poincar\'{e} supergravity and the corresponding counting of states, introduces auxiliary fields, which fall in three classes. In any case the total number of off-shell degrees of freedom is $40_B+40_F$ of which $24_B+24_F$ are included in the Weyl multiplet and are not interesting since they give rise to a ghost spin-2 long multiplet when the $Weyl^2$ term is added to the Einstein one. The remaining $16_B+16_F$ states correspond to the $R+R^2$  theory and should then be physical and dual to
a standard matter-coupled $\mathcal{N}=2$ supergravity, whose matter content corresponds to the $32$ states. In \cite{deWit:1979pq} an analysis of the (linearized) quadratic approximation of a fourth-derivative supergravity theory was performed and the spectrum, excluding the spin-2 ghost multiplet, was found to fall in two massive $8+8$ long vector multiplets, then giving the remaining degrees of freedom to complete the total number of off-shell states of $\mathcal{N}=2$ Poincar\'{e} supergravity. In one of the three different off-shell formulations, the massive states feel a massive vector and two massive hypermultiplets. However such an assignment cannot correspond to $R+R^2$ supergravity since it requires the gauging of the central charge. From the previous arguments we are led to conclude that, using any of the other two formulations, $R+R^2$ supergravity, if it exists, must be dual to a standard  $\mathcal{N}=2$ supergravity coupled to two long massive vector multiplets. Each multiplet then contains a spin $1$, four spin $1/2$ and five scalar fields, and the $\mathcal{N}=2$  analogue of the St\"{u}ckelberg formulation of massive vector states should be based on a gauging of two abelian isometries of a two-dimensional quaternionic K\"{a}hler manifold \cite{Bagger:1983tt}, which then provides the supersymmetric Higgs mechanism. This is the gravity analogue of the $\mathcal{N}=2$ Higgs effect, which was first studied by Fayet at the birth of $\mathcal{N}=2$ supersymmetry \cite{Fayet:1975yi}. The two massive vector multiplets correspond at the linearized level to two different  $\mathcal{N}=2$  higher-derivative invariants. We could consider the effective theory resulting from sending the mass of one of them to infinity. It should contain a single massive vector multiplet. In this paper we also analyze possible cosmological potentials that follow from this scenario.

 Note that in this massive theory the vector multiplet should have non-vanishing mass for all vanishing values of the scalar fields, which can only occur if the corresponding isometries are non-compact. Non-compact isometries fall into two classes: \emph{hyperbolic} and \emph{parabolic}, in contrast to the compact \emph{elliptic} ${\rm U}(1)$-isometries \cite{Ferrara:2013eqa,Ferrara:2014rya}. The different nature of these isometries will play a major role  for the search of $\mathcal{N}=2$ potentials possibly describing an inflaton.\footnote{For recent works on inflation in the context of gauged $\mathcal{N}=2$ theory see also \cite{Fre:2013tya}.} An important requirement, in analogy to the $\mathcal{N}=1$ case, is that the theory should have, at some finite distance point in the moduli space, an unbroken $\mathcal{N}=2$ Minkowski vacuum. A similar property of the Starobinsky potential $V=g^2\,\left(1-\rme^{-\sqrt{\frac{2}{3}}\phi}\right)^2$ is that it vanishes for $\phi=0$. Motivated by these features, in this paper we analyze a large class of suitable $\mathcal{N}=2$ theories which have two massive abelian vector multiplets coupled to gravity. In analogy with the $\mathcal{N}=1$ theory, it is tempting to infer that the scalaron will be one of the five scalars of one of the two massive vector multiplets, while the role of the other multiplet would be to contain the two goldstinos of the $\mathcal{N}=2$ supersymmetry breaking. The two goldstinos would then be part of the spin $1/2$ massive states that are the fermionic partners of the  $\mathcal{N}=2$ massless vector decomposition.
An important restriction comes if we require the hypermultiplet manifold to embed in a natural way the scalaron field. This restriction uniquely selects the exceptional quaternionic space ${\rm G}_{2(2)}/{\rm SU}(2)^2$ which, having rank 2, contains two dilaton-like scalars and six axions, one of which becomes the longitudinal mode of the vector.
The outcome of this search is surprising  since we find that the only commuting abelian non-compact isometries, compatible with the required existence of  an $\mathcal{N}=2$ Minkowski vacuum, are of  hyperbolic type. The corresponding potentials typically diverge for large values of $\phi$ instead of exhibiting  a plateau. Conversely we find theories with Starobinsky-like potentials \cite{Kallosh:2013yoa} $V=g^2\,\left(1-\rme^{-\sqrt{\frac{2}{3\alpha}}\phi}\right)^2$ in the one-multiplet case, with $\alpha=1/3$ and $2/3$. Moreover, in the two vector-multiplet case, we find a model with $\alpha=1$ (Starobinsky potential) but where one of the two
vector fields has vanishing mass, so that the theory contains a massless vector and a massless hypermultiplet. As we show in Sect. \ref{parabola}, the above $\alpha$-deformed potentials can only originate, in the class of $\mathcal{N}=2$ models considered here, from the gauging of non-semisimple quaternionic isometries (to be dubbed \emph{parabolic}). In the context of $\mathcal{N}=1$ supergravity this is consistent with the results of \cite{Ferrara:2013eqa,Ferrara:2014rya}. Group theory then constrains the values of $\alpha$ to be of order one for ${\cal N}=2$, in apparent contrast with the results of BICEP2 which seem to require $\alpha$ to be, at least,  of order 50.
\par
The paper is organized as follows. \\
In Section \ref{ss:multipletsN2} we give the multiplets of  $\mathcal{N}=2$ conformal supergravity, with the different off-shell formulations.\\
In Section \ref{ss:N2gauging} we discuss the various models, both in the presence of one or two vector-multiplets + hypermultiplets. The search is effected, in the single vector multiplet + hypermultiplet case,  by allowing all possible special K\"{a}hler and quaternionic manifolds that are symmetric spaces. In this case there are two possible choices of the special K\"{a}hler and quaternionic- K\"{a}hler spaces, respectively. When the former is of type $\mathbb{CP}^1$, the Starobinsky-like potentials are found upon truncation to the quaternionic scalars. In fact, such truncation, as  we prove on general grounds, is consistent only in the  $\mathbb{CP}^1$  special K\"{a}hler geometry, by virtue of the fact that for this space, the holomorphic rank-3 tensor $C_{ijk}$ vanishes.
As far as the two vector multiplet case is concerned, we find an $\alpha=1$ potential upon gauging of only one parabolic isometry of $\mathfrak{g}_{2(2)}$.

\section{Multiplets and off-shell fields for \texorpdfstring{${\cal N}=2$}{N=2}}
\label{ss:multipletsN2}
The experience of pure gravity and pure ${\cal N}=1$ supergravity tells us that the off-shell fields of Poincar\'{e} gravity describe the physical fields of theories dual to (super)Poincar\'{e} with curvature square terms in the action. For pure gravity, omitting total derivatives, we write
\begin{equation}
S=\int \rmd^4 x\,\frac{1}{2\kappa^2}\sqrt{g}  \left[ R + \alpha R^2 + \beta (R_{\mu \nu }R^{\mu \nu } -\ft13 R^2)\right]\,.
\label{SR2bos}
\end{equation}
The first term describes the massless graviton field, using the field $e_\mu ^a$, which has 6 off-shell degrees of freedom (dof), counting 16 components minus the 10 gauge dof of the Poincar\'{e} algebra. One can describe these also as 5 dof of a conformal frame field\footnote{Subtracting the dilation gauge degrees of freedom. The special conformal degrees of freedom eliminate the 4 components of the gauge field of dilatations, $b_\mu $.}, and a scalar compensator field, $\phi $, whose value determines the gravitational constant $\kappa $. When adding the $\alpha $-term, the theory describes also a massive physical scalar dof with $m_0{}^{-2}=12\alpha \kappa ^2$, while the $\beta $-term leads to a ghost spin 2 with $m_2{}^{-2}=-2\beta \kappa ^2$  \cite{Stelle:1977ry}. In the dualized theory the physical scalar field is the `auxiliary' component $\phi $. The massive spin 2 ghost can be identified with the conformal part of $e_\mu ^a$. In this way the 6 off-shell dof from the Poincar\'{e} theory describe the massive states in the action (\ref{SR2bos}). How this is done in practice is shown in Appendix \ref{app:confdual}.

The same mechanism works also for ${\cal N}=1$: the off-shell dof of the super-Poincar\'{e} theory can be written as a Weyl multiplet, the analogue of the conformal $e_\mu {}^a$ in the bosonic theory, and a compensating multiplet. The curvature squared terms describe a massive ghost multiplet (the analogues of the $\beta $-term) using the components of the Weyl multiplet, and a physical massive multiplet (the analogues of the $\alpha $-term) built from the fields in the compensating multiplets.

We will now investigate how the auxiliary fields of ${\cal N}=2$ supergravity can fulfill the same role. To understand how they can fit in massive multiplets, we first repeat the content of the massive representations (without central charge) of ${\cal N}=2$ supersymmetry \cite{Ferrara:1980bh}.

\subsection{Massive multiplets}
The spin 2 ghost long  multiplet contains (using  $\USp(4)$ representations)
\begin{equation}
  \begin{array}{|c|cccc|}\hline
    \mbox{spin} & \mbox{\#} & \mbox{dof/particle}& \mbox{bosonic} & \mbox{fermionic} \\ \hline
    2 & 1 & 5& 5 &   \\
    3/2 & 4 & 4&   & 16 \\
    1 & 5+1 & 3& 18 &   \\
    1/2 & 4 & 2&   & 8 \\
    0 & 1 & 1& 1 &   \\ \hline
    TOTAL &   &  & 24 & 24 \\ \hline
  \end{array}
 \label{spin2ghostm}
\end{equation}
The massive spin 1 long multiplet is
\begin{equation}
  \begin{array}{|c|cccc|}\hline
    \mbox{spin} & \mbox{\#} & \mbox{dof/particle}& \mbox{bosonic} & \mbox{fermionic} \\ \hline
    1 & 1 & 3& 3 &   \\
    1/2 & 4 & 2&   & 8 \\
    0 & 5 & 1& 5 &   \\ \hline
    TOTAL &   &  & 8 & 8 \\ \hline
  \end{array}
 \label{massivevm}
\end{equation}
The massive spin 1/2 multiplet is
\begin{equation}
  \begin{array}{|c|cccc|}\hline
    \mbox{spin} & \mbox{\#} & \mbox{dof/particle}& \mbox{bosonic} & \mbox{fermionic} \\ \hline
    1/2 & 2 & 2&   & 4 \\
    0 & 4 & 1& 4 &   \\ \hline
    TOTAL &   &  & 4 & 4 \\ \hline
  \end{array}
 \label{massivehm}
\end{equation}
\subsection{Field representations}

Off-shell Poincar\'{e} supergravities are built from the Weyl multiplet and two compensating multiplets \cite{deWit:1981tn}. For the different off-shell formulations that are used to build Poincar\'{e} actions, the first compensating multiplet is always a gauge multiplet. The second one can be either a so-called non-linear multiplet (version I, \cite{deWit:1979pq,Fradkin:1979as,deWit:1980ug}), a hypermultiplet (version II \cite{deWit:1981tn}) or a tensor multiplet (version III \cite{deWit:1982na}). We now consider the fields of these multiplets and assign them to massive states.

The Weyl multiplet has the following fields, which represent the components of a
massive spin 2 multiplet (\ref{spin2ghostm})

\begin{equation}
\begin{array}{|c|c|ccccc|}\hline
 \mbox{field}& \mbox{dof} & \mbox{spin } 2&\mbox{spin } \ft32&\mbox{spin } 1&\mbox{spin } \ft12&\mbox{spin } 0   \\
\hline
e_\mu {}^a & 5 & 1 &&&&\\
b_\mu & 0 &  &&&& \\
V_{\mu i}{}^j & 9 & &&3&&   \\
A_\mu & 3 & &&1&&  \\
T^-_{ab} & 6 && &2&&\\
D & 1 & &&&& 1 \\
\psi _\mu {}^i & 16 & &4&&&\\[1mm]
\chi ^i & 8 & &&&4& \\ \hline
\end{array}     \label{tblWeylN2}
\end{equation}
In this and the following tables, the off-shell number of degrees of freedom (dof) are given, subtracting all the superconformal gauge dof. Then we give the massive spin representations to which the fields correspond.

The off-shell gauge multiplet, which we need for the `minimal off-shell representation' (which does not allow a Poincar\'{e} action) represents a massive spin 1 multiplet (\ref{massivevm}):
\begin{equation}
\begin{array}{|c|c|ccc|}\hline
 \mbox{field}& \mbox{dof} & \mbox{spin } 1&\mbox{spin } \ft12&\mbox{spin } 0 \\
\hline
X & 2 & & & 2\\
A_\mu & 3 & 1 &&  \\
\vec{Y}& 3 & & & 3\\
\Omega_i  & 8 &   & 4 &\\ \hline
\end{array}
\end{equation}
The off-shell hypermultiplet represents 2 massive spin 1/2 multiplets (\ref{massivehm}):
\begin{equation}
\begin{array}{|c|c|cc|}\hline
 \mbox{field}& \mbox{dof} & \mbox{spin } \ft12&\mbox{spin } 0   \\
\hline
q^X & 4 & & 4\\
F^X & 4 & & 4  \\
\zeta ^A  & 8 &   4 &\\ \hline
\end{array}
\end{equation}
The  tensor multiplet represents a massive spin 1 multiplet (\ref{massivevm}):
\begin{equation}
\begin{array}{|c|c|ccc|}\hline
 \mbox{field}& \mbox{dof} & \mbox{spin } 1&\mbox{spin } \ft12&\mbox{spin } 0  \\
\hline
L_{ij} & 3 & & & 3\\
E_{\mu \nu }& 3 & 1&&\\
G &   2 & & & 2  \\
\varphi _i & 8 &  & 4 &\\ \hline
\end{array}
\end{equation}
and the same holds for the non-linear multiplet :
\begin{equation}
\begin{array}{|c|c|ccc|}\hline
 \mbox{field}& \mbox{dof} & \mbox{spin } 1&\mbox{spin } \ft12&\mbox{spin } 0  \\
\hline
\Phi _i^\alpha  & 3 & && 3\\
M_{[ij]} & 2 & & & 2\\
V_a &   3 & 1 & &  \\
\lambda  _i & 8 & & 4& \\ \hline
\end{array}
\end{equation}

As a representation of massive fields, the Weyl multiplet and the gauge multiplet thus lead to the
fields of massive spin 2 + spin 1 multiplets. The three
40+40 sets of auxiliary fields mentioned above give a further spin 1 massive multiplet in the first and third cases, and two spin 1/2 multiplets in the second case.

Therefore, following the scheme that we learned from the ${\cal N}=0$ and ${\cal N}=1$ theories, the off-shell field content of ${\cal N}=2$ Poincar\'{e} supergravity provides fields for the following states in an $R+R^2$ action:
\begin{equation}
   4+4 \mbox{ (massless sugra) }+ 24+24\mbox{ (massive spin 2 ghost) }+2(8+8) \mbox{ (2 massive physical) }
 \label{N2states}
\end{equation}
The massive states correspond to the decomposition of the 40+40 off-shell dof of the super-Poincar\'{e} theory in Weyl + 2 compensating multiplets. With
the non-linear or the tensor multiplet compensator the last two multiplets are two spin 1 multiplets. With the hypermultiplet compensator these are a
spin 1 and two spin 1/2 multiplets.

\subsection{Structure of the first set of auxiliary fields}
We summarize now where to find the relevant formulae for the first set of
auxiliary fields, as found in \cite{deWit:1979pq}. In Table \ref{splitfields}
we list the fields, their number of dof, and how they are split in the conformal setting and rearranged in the multiplets.
%XXX This table should still be checked what concerns the middle column: how the fields connect XXX
\begin{table}[htbp]
  \caption{\it Fields of the first off-shell formulation of supergravity and their appearance in conformal multiplets}
  \label{splitfields}
\begin{center}
\[   \begin{array}{|cc|cc|ccc|} \hline
   \mbox{field}& \mbox{dof}& \mbox{split}& \mbox{type}
   &\mbox{Weyl }& \mbox{vector}& \mbox{non-lin.}\\ \hline
    e_\mu ^a & 6 & 5 &   & e_\mu ^a &   &   \\
      &   & 1 & %D-\ft12 R :
      \mbox{ dilaton} &   & \Re X &   \\
    A_\mu  & 4 & 3 & \U(1)\mbox{ gauge f.} & A_\mu  &   &   \\
     &   & 1 & %\partial \cdot A:\
     \U(1) \mbox{ compens.}  &   & \Im X &   \\
    {\cal V}_\mu {}^i{}_j & 12 & 9 & \SU(2)\mbox{ gauge f.}  & {\cal V}_\mu {}^i{}_j  &   &   \\
      &   &  3 & %\partial \cdot {\cal V}_\mu {}^i{}_j:\
      \SU(2) \mbox{ compens.}  &   &   & \Phi _i^\alpha  \\
    V_\mu  & 4 & 3 & %\partial \cdot V -D-\ft13 R=0
    &   &   & V_\mu  \\
      &   & 1 & %D=\partial \cdot V-\ft13 R
      & D &   &   \\
    {\cal M}^{[ij]} & 2 & 2 & %M^{[ij]}
    &   &   &  {\cal M}^{[ij]} \\
    {\cal S}^{(ij)} & 3 & 3 & %Y^{ij}
    &   & Y^{ij} &   \\
    t_{\mu \nu }^{[ij]} & 6 & 6& %T_{\mu \nu }^{[ij]}
    & T_{\mu \nu }^{[ij]}  &   &      \\
    B_\mu {}^{[ij]} & 3 &3& % B_\mu
      &   & B_\mu  &   \\ \hline
    \psi _\mu ^i & 24 & 16 & Q\mbox{ gauge f.} & \psi _\mu ^i  &   &   \\
      &   & 8 & %\gamma \cdot R:\
      Q \mbox{ compens.} &   & \xi ^i &   \\
    \xi ^i & 8 & 8 & %\chi ^i-\ft13\gamma \cdot R
    & \chi ^i &   &   \\
    \lambda ^i & 8 & 8 &  &   &   & \lambda ^i \\   \hline
   \mbox{TOT} & 40+40 & 40+40 &   & 24+24 & 8+8 & 8+8    \\   \hline
  \end{array}
 \]
\end{center}
\end{table}
Some fields of \cite{deWit:1979pq} have been rewritten in complex combinations, like ${\cal V}_{\mu \,i}{}^j=V_\mu ^{[ij]}-\rmi A_\mu ^{(ij)_S}$, ${\cal S}^{(ij)}=S\delta ^{ij}-\rmi P^{(ij)_S}$,
${\cal M}^{[ij]}=M^{[ij]}+\rmi N^{[ij]}$.
The indication $(ij)_S$ means symmetric traceless. The
triplet notations follow the
conventions in Appendix 20A in \cite{Freedman:2012zz}.

%\subsubsection{Weyl multiplet}
%The linearized Lagrangian is in (2.13) of \cite{deWit:1979pq}. The full action
%is in (5.18) of \cite{Bergshoeff:1981is}. The redefinition from the fields in
%(\ref{fieldsN2Poinc}) to the conformal fields is in (2.8) and (3.6) of
%\cite{deWit:1980ug}, thus for the linear parts this is
%\begin{eqnarray}
%  T_{ab}^{[ij]}&=&(t-\sqrt{2}F(B))_{\mu \nu }^{[ij]}\,,\nonumber\\
%  \chi ^i&=&(\chi -\ft13\gamma \cdot R^P)^i\,,\nonumber\\
%  D&=&\partial \cdot V-\ft13\kappa ^{-1}R^P+\ldots \,.
% \label{conffromPoinc}
%\end{eqnarray}
%The conformal transformations are in (3.7) of \cite{deWit:1980ug}.
%Therefore the bosonic components of the linearized part of the Weyl-curvature multiplet are:
%\begin{align}
%&C_{\mu\nu\rho\sigma}\,\, (\mbox{\small Weyl tensor})\,;\,\,\,F(\mathcal{V}_i{}^j)_{\mu\nu},\,F(A)_{\mu\nu}\,\, (\mbox{\small ${\rm U}(2)$ gauge field-strengths})\,\,;\,\,\,\partial^\mu T_{\mu\nu}^{[ij]}\,\,;\,\,\,\partial^\mu \tilde{T}_{\mu\nu}^{[ij]}\,;\,\,D\,.
%\end{align}
%When the Lagrangian (2.13) is added to the Einstein term, one gets a massive spin-2 ghost multiplet, where the 6 massive vectors come from the four ${\rm  SU}(2)\times {\rm U}(1)$ gauge fields and an extra ${\rm U}(1)$ doublet obtained from  the dual and anti self-dual combinations of $T_{\mu\nu}^{[ij]}$ and $\tilde{T}_{\mu\nu}^{[ij]}$.

\subsection{Actions for \texorpdfstring{${\cal N}=2$}{N=2}}
The off-shell actions for several parts have been written down in \cite{deWit:1979pq}. See especially equations (2.1) for Poincar\'{e} supergravity,
(2.9) for a tensor multiplet, (2.12) for a vector multiplet, and (2.13) is the conformal Weyl square action. The construction there is equivalent to
conformal supergravity with a compensating vector multiplet and a tensor multiplet.\footnote{Actually this tensor multiplet is the linear part of
what has been called later \cite{deWit:1981tn} the non-linear multiplet.} Hence it is a massive spin 2 multiplet and 2 massive spin 1 multiplets.
Thus, e.g. there are 8 massive vectors.

In \cite{deWit:1980ug} these fields appear in a different way. Essentially the fields that are in Table~\ref{splitfields} in the vector multiplet, appear
in a tensor multiplet, while those of the non-linear multiplet appear in a vector multiplet. The recombination of the fields of the Poincar\'{e} theory
in these can be found in \cite{deWit:1980ug}, especially (3.6) for the Weyl multiplet, (5.8) for the vector multiplet and (6.1) for the tensor
multiplet.

These are the actions for the kinetic multiplets of the conformal compensating multiplets, as we will show below.

\subsubsection{Tensor calculus rules}

Suppose $\Phi $ is chiral multiplet. We further use 'the constraint' for
\begin{equation}
  D^iD^j \Phi =\varepsilon ^{ik}\varepsilon ^{j\ell }\bar D_k\bar D_\ell\bar \Phi\,.
 \label{constraintV}
\end{equation}
If $\Phi $ satisfies the constraint, we write $\Phi _C=0$, and we call it a vector multiplet $\Phi =V$ and $V_C=0$.

A general chiral multiplet can be written as
\begin{equation}
  \Phi =V+L\,,
 \label{PhiVL}
\end{equation}
where $V$ and $L$ are both chiral, and $V$ satisfies the constraint $V_C=0$. Then $L$ is a linear multiplet. In other words: a linear multiplet is a
chiral multiplet modulo a vector multiplet.

Note that $\rmi V $ does not satisfy the constraint. But we can write
\begin{equation}
  \rmi V=W + T(V)\,,\qquad W_C=0\,.
 \label{decompiV}
\end{equation}
$T(V)$ is thus $\rmi V$ modulo a vector multiplet, hence it is a linear multiplet.

Now about actions. The actions are built from the real part of the highest
component of a chiral multiplet, which we indicate as $[$chiral$]_C$. An
important property is that for two vector multiplets $V$ and $V'$
\begin{equation}
 [\rmi V V']_C=0 \,.
 \label{VV'C0}
\end{equation}
Note however that $[ VV']_C\neq 0$. In fact, using first (\ref{decompiV}) and
then (\ref{VV'C0}):
\begin{equation}
  [ V\,V']_C=[-\rmi(W + T(V))\, V']_C= [-\rmi T(V)\, V']_C\,.
 \label{kinVM}
\end{equation}
For $V'=V$, this is the quadratic kinetic action of a vector multiplet.

In general we can consider this as an action built from a product of a vector
and a linear multiplet, considering $T(V)$ as an arbitrary linear multiplet.
Hence we have an action symbolically
\begin{equation}
  S= L\times V\,,
 \label{LtimesV}
\end{equation}
such that $T(V)\times V$ is the same as $[VV]_C$. This action formula is given
in (2.5) in \cite{deWit:1982na}.

There is a construction to build a vector multiplet from a linear multiplet,
$K(L )$ that in rigid supersymmetry takes the highest component of a linear
multiplet, but is, due to the chiral weights, more involved in the conformal
framework. For the kinetic action of a linear multiplet we use
\begin{equation}
%  [K(\Phi )\,\Phi]_C= [K(L)\,\Phi ]_C= [K(L)\,L]_C \,.
L\times K(L)\,.
 \label{kinLinear}
\end{equation}

\subsubsection{Application for massive actions}

We symbolically write
\begin{equation}
  \poin = \mbox{Weyl} + V_0+ L_0\,,
 \label{Poincfromconf}
\end{equation}
where $\poin$ is the Poincar\'{e} multiplet, and $V_0$ and $L_0$ are the compensating vector multiplets and linear multiplets. The Poincar\'{e} action is
then built from the kinetic action of those:
\begin{equation}
  L_R= T(V_0)\times V_0+ L_0\times K(L_0)\,.
 \label{Poincareaction}
\end{equation}
The first term can be written as $[V_0V_0]_C$.

To build higher derivative actions we add
\begin{equation}
  L_M=\frac{1}{M_1^2}T(V_0)\times K(T( V_0))+ \frac{1}{M_2^2}T(K(L_0))\times  K(L_0)\,.
 \label{LMsymbolisch}
\end{equation}
It is instructive to recall the linearized bosonic components of the vector and linear multiplets $V_0$ and $L_0$. $V_0$ (or rather its field-strength multiplet) has the following bosonic components
\begin{align}
X&= \rmi\,\partial\cdot A+\left(\partial\cdot V-\frac{1}{2}\, R\right)\,\,,\,\,\,Y^{ij}={\cal S}^{ij}\,\,,\,\,\,\mathcal{T}_{\mu\nu}{}^{ij}= (-t+\frac{1}{\sqrt{2}}\,F(B))_{\mu \nu }^{[ij]}\,.
\end{align}
The massive multiplet generated by $V_0$ is the one which contains the $R^2$- term.\par
The second massive multiplet is generated by the linear multiplet whose bosonic components are:
\begin{align}
& F(V)\,\,\,,\,\,\,\,{\cal M}^{ij}\,\,,\,\,\partial\cdot \mathcal{V}^i{}_j\,.
\end{align}
The linearized quadratic action given in (\ref{LMsymbolisch}) gives, when added to the Einstein term,  therefore two massive long vector multiplets with masses $M_1$ and $M_2$.
\section{\texorpdfstring{$\mathcal{N}=2$}{N=2} gauging}
\label{ss:N2gauging}
We study gaugings in the $\mathcal{N}=2$ model with the following symmetric scalar manifold:
\begin{align}
\mathcal{M}_{scal}&=\mathcal{M}_{SK}\times \mathcal{M}_{QK} \,,\nonumber\\
\mathcal{M}_{SK}&=\frac{{\rm SU}(2,1)}{{\rm U(2)}}\,\,;\,\,\,\mathcal{M}_{QK}=\frac{{\rm G}_{2(2)}}{{\rm SU(2)\times SU(2)}}\,,
\end{align}
under the following conditions:
\begin{itemize}
\item[a]{Existence of an $\mathcal{N}=2$ Minkowski vacuum;}
\item[b]{Two massive vector multiplets on the vacuum.}
\end{itemize}
To this end we gauge two commuting isometries of $\mathfrak{g}_{2(2)}$.
This requires choosing a two-dimensional abelian gauge group inside ${\rm G}_{2(2)}$. In order to make the analysis independent of the initial symplectic frame, it is useful to describe  the gauge algebra generators by the (redundant) symplectic  notation $X_M=(X_\Lambda,\,X^\Lambda)=(X_0,X_1,X_2,X^0,X^1,X^2)$, see \cite{deWit:2005ub}. Of the six $X_M$, only two are independent generators. We can expand $X_M$ in the $\mathfrak{g}_{2(2)}$ generators $t_\alpha$ (for which we use the conventions in Appendix \ref{g2}) through the embedding tensor $\theta_M{}^\alpha$  \cite{Cordaro:1998tx,Nicolai:2001sv,deWit:2002vt,deWit:2005ub,deWit:2005ub}
\begin{equation}
X_M=\theta_M{}^\alpha \,t_\alpha\,.
\end{equation}
The embedding tensor $\theta_M{}^\alpha$ is subject to the locality requirement  and a further condition coming from the closure of the abelian algebra:
\begin{equation}
\theta_\Lambda{}^\alpha\theta^{\Lambda\,\beta}-\theta_\Lambda{}^\beta\theta^{\Lambda\,\alpha}=0\,\,;\,\,\,[X_M,\,X_N]=0\,.
\end{equation}
The Lagrangian density reads ($\kappa^2=c=\hbar=1$):
\begin{align}
e^{-1}\mathcal{L}=\frac{R}{2}-h_{uv}\partial_\mu q^u\,\partial^\mu q^v-g_{i\bar{\jmath}}\partial_\mu z^i\,\partial^\mu \bar{z}^{\bar{\jmath}}-V\,,\label{lagra}
\end{align}
where $q^u$ are the quaternionic scalars, $z^i$ the complex vector-multiplet scalars. The scalar potential has the general form
given in \cite{Andrianopoli:1996vr}, which can be modified in the following symplectic invariant fashion \cite{Dall'Agata:2003yr,D'Auria:2004yi,deWit:2005ub,Louis:2010ui,deWit:2011gk,Louis:2012ux}:
\begin{align}
V&=(4\,k_M^u\,k_N^v\,h_{uv}+k_M^i\,k_N^{\bar{\jmath}}\,g_{i\bar{\jmath}})\,\overline{V}^M\,V^N+(g^{i\bar{\jmath}}\,D_iV^M\,D_{\bar{\jmath}}\overline{V}^N-3\,\overline{V}^M\,V^N)\,\mathcal{P}_N^x\,\mathcal{P}_M^x\,.
%
%\mathcal{V}_{hyperino}+\mathcal{V}_{gaugino,1}+\mathcal{V}_{gaugino,2}+\mathcal{V}_{gravitino}\,,\nonumber\\
%\mathcal{V}_{hyperino}&= 4\,\overline{V}^M\,V^N\,\,k_M^u\,k_N^v\,h_{uv}\,\,\,;\,\,\,\,
%\mathcal{V}_{gaugino,1}= \overline{V}^M\,V^N\,k_M^i\,k_N^{\bar{\jmath}}\,g_{i\bar{\jmath}}\,,\nonumber\\
%\mathcal{V}_{gaugino,2}&= g^{i\bar{\jmath}}\,D_iV^M\,D_{\bar{\jmath}}\overline{V}^N\,\mathcal{P}_M^x\,\mathcal{P}_N^x\,\,;\,\,\,
%\mathcal{V}_{gravitino}=-3\,\overline{V}^M\,V^N\,\mathcal{P}_N^x\,\mathcal{P}_M^x\,,
\end{align}
where $V^M=(L^\Lambda,\,M_\Lambda)$ is the section of the symplectic ${\rm U}(1)$-bundle over the special K\"{a}hler manifold. In our case, $k_M{}^i=0$ while the gauge Killing vectors and the corresponding momentum maps can be written in the form:
\begin{equation}
k_M^u=\theta_M{}^\alpha\,k_\alpha{}^u\,\,;\,\,\,\mathcal{P}_M^x=\theta_M{}^\alpha\,\mathcal{P}_\alpha{}^x\,,
\end{equation}
$k_\alpha{}^u$ and $\mathcal{P}_\alpha{}^x$, $x=1,2,3$,  being the Killing vector and the momentum map corresponding to the isometry $t_\alpha$.\par
At the $\mathcal{N}=2$ bosonic background $z^i=z^i_0,\,q^u=q^u_0$, the Killing spinor equations:
\begin{align}
\delta \psi_{A\mu}&= -\frac{1}{2}\,(\sigma_x)_{AB}V^M\,\mathcal{P}_M^x\,\gamma_\mu\,\epsilon^B=0\,\,;\,\,\,\delta\lambda^{iA}=
i\,(\sigma_x)^{AB}\bar{D}^i\overline{V}^M\,\mathcal{P}_M^x\,\epsilon_B=0\,,\nonumber\\
\delta\zeta_{\hat{\alpha}}
&=2\,\mathcal{U}^A_{\hat{\alpha}\,u}\,\overline{V}^M\,k_M^u\,\epsilon_A=0\,\,\,\,\,\,\,\forall\epsilon^A\,,
\end{align}
imply, at the vacuum:
\begin{equation}
\mathcal{P}_M^x=\theta_M{}^\alpha\,\mathcal{P}_\alpha{}^x=0\,,\qquad V^M\,k_M^u=0\,.
\end{equation}
Since we refrain from gauging the central charge, we require that, for $z^i=z_0^i$, the stronger condition $V^M\,\theta_M{}^\alpha=0$ holds.

We look for models in which the inflaton is one of the hyper-scalars. This constrains our models by the requirement that the restriction to the dilatonic scalars of the quaternionic K\"{a}hler coset be consistent.
In particular it should be consistent to fix the scalars $z^i$ to fixed at their background values $z_0^i$. If this is the case, the potential  effectively  reduces to:
\begin{equation}
V(q^u,\theta)=\left(g^{i\bar{\jmath}}\,D_iV^M\,D_{\bar{\jmath}}\overline{V}^N\right)_{z^i=z^i_0}\,\mathcal{P}_M^x\,\mathcal{P}_N^x\,.
\end{equation}
However fixing $z^i=z^i_0$ and leaving $q^u$ free is not always a consistent truncation. Indeed one can prove that
\begin{align}
\left.\partial_iV\right\vert_{z^i=z^i_0}&=\left(\overline{V}^M\,D_iV^N\right)_{z^i=z^i_0}\,(4\,k_M^u\,k_N^v\,h_{uv}-3 \mathcal{P}_N^x\,\mathcal{P}_M^x)+\nonumber\\
&+\left(g^{j\bar{\jmath}}\,D_iD_jV^M\,D_{\bar{\jmath}}\overline{V}^N+g^{j\bar{\jmath}}\,D_jV^M\,D_iD_{\bar{\jmath}}\overline{V}^N\right)_{z^i=z^i_0}\,\mathcal{P}_M^x\,\mathcal{P}_N^x=\nonumber\\
&=\left(i\,g^{j\bar{\jmath}}\,C_{ijk}D^k\overline{V}^M\,D_{\bar{\jmath}}\overline{V}^N+D_iV^M\,\overline{V}^N\right)_{z^i=z^i_0}\,\mathcal{P}_M^x\,\mathcal{P}_N^x=\nonumber\\
&=\left(i\,C_{ijk}D^j\overline{V}^M\,D^k\overline{V}^N\right)_{z^i=z^i_0}\,\mathcal{P}_M^x\,\mathcal{P}_N^x\,.
\end{align}
where we have used the properties $V^M\,k_M{}^u=0$ and $V^M\,\mathcal{P}_M^x=0$ at $z^i=z_0^i$. Clearly the non-vanishing tensor $C_{ijk}$ can be an obstruction to the truncability of the model to the hyperscalars, i.e. the consistency of the truncation to the hyperscalars implies a condition on the gauging. We can therefore truncate the model to the hyperscalars with no constraint on the gauge group when $\mathcal{M}_{SK}$ is $\mathbb{CP}^n$ (minimal coupling).\par
Since the condition $V^M(z^i_0,\bar{z}_0^{\bar{\imath}})\,\theta_M{}^\alpha=0$ is independent of the symplectic frame, we can choose the special coordinate one with prepotential
\begin{equation}
\mathcal{F}(z^i)=-\frac{\rmi}{2}\,\left(1-(z^1)^2-(z^2)^2\right)\,.
\end{equation}
We can also fix the special K\"{a}hler manifold isometries by choosing  $z_0^i=0$, so that we must choose $\theta^{0\alpha}=\theta_0{}^\alpha=0$. The potential becomes:
 \begin{equation}
V(q^u,\theta)= \left(g^{i\bar{\jmath}}\,D_iV^M\,D_{\bar{\jmath}}\overline{V}^N\right)_{z^i=0}\mathcal{P}_M^x\,\mathcal{P}_N^x=
\frac{1}{2}\left(\mathcal{P}_a^x\,\mathcal{P}_a^x+\mathcal{P}^{a\,x}\,\mathcal{P}^{a\,x}\right)\,\,\,,\,\,\,a=1,2\,.
 \end{equation}
Using the locality condition on the embedding tensor and the ${\rm U}(2)$ isotropy group of the special K\"{a}hler manifold, the components of the embedding tensor can be significantly reduced (we can for instance choose, with no loss of generality, as non-vanishing entries of $\theta_M{}^\alpha$: $\theta_1{}^\alpha,\,\theta_2{}^\alpha$). Since however we restrict to the origin of the special K\"{a}hler manifold, we have verified that the outcome of our analysis does not depend on the initial choice of the symplectic frame and we shall therefore illustrate, for the sake of simplicity, our results in the frame where the only non-vanishing components are $\theta_1^{\alpha},\,\theta_{2}{}^\alpha$.  The scalar potential then reads:
  \begin{equation}
V(q^u,\theta)= \frac{1}{2}\sum_{x=1}^3\left[(\mathcal{P}_1^x)^2+(\mathcal{P}_2^{x})^2\right]\,.
 \end{equation}
 \par
 The two dilatons in the quaternionic manifold are denoted by $U,\,\varphi$, the latter defines the imaginary part of the complex coordinate $T=y-\rmi\,\rme^{\varphi}$ spanning a cubic $T^3$-submanifold of the quaternionic one.
 The scalar Lagrangian density, truncated to the dilatons, reads:
\begin{equation}
e^{-1}\mathcal{L}=\frac{R}{2}-\partial_\mu U\partial^\mu U-\frac{3}{4}\,\partial_\mu \varphi\partial^\mu \varphi-V\,.
\end{equation}
By virtue of the homogeneity of the quaternionic K\"{a}hler manifold, to any $\mathcal{N}=2$ extremum $q_0^u$ of $V(q^u,\theta)$ there corresponds a gauging where the same vacuum is at the origin $q^u=0$.\footnote{In \cite{Dibitetto:2011gm,DallAgata:2011aa,Dall'Agata:2012bb} this property has been used to define general approach for finding new  vacua in $\mathcal{N}=4$ and $8$ theories.} Defining the origin as the point where the coset representative $\mathbb{L}(q^u)$ of the quaternionic manifold is the identity, see Appendix \ref{g2}, the embedding tensor $\theta'_M{}^\alpha$ associated with the new gauging is obtained by a ${\rm G}_{2(2)}$-transformation of the original one $\theta_M{}^\alpha$: $\theta'_M{}^\alpha=\theta_M{}^\beta\,\mathbb{L}(q_0)_\beta{}^\alpha$. We can therefore look for $\mathcal{N}=2$ vacua at the origin, provided we choose $X_1,\,X_2$ generic in $\mathfrak{g}_{2(2)}$. \par
The condition of $\mathcal{N}=2$ supersymmetry at the origin $z^i=0,\,q^u=0$  are therefore, in the chosen symplectic frame:
\begin{align}
V^M(0)\,\theta_M{}^\alpha &=0\,\,\Leftrightarrow\,\,\,\,\theta^{0\alpha}=\theta_0{}^\alpha=0\,,\nonumber\\
\mathcal{P}_M{}^x(q^u=0)&=-\frac{1}{2}\,{\rm Tr} (J^x\,X_M)=0\,.\label{n2}
\end{align}
The last condition amounts to requiring the gauge generators $X_1,\,X_2$ to be orthogonal to the $\mathfrak{su}(2)_R$ algebra of the quaternionic structure, generated by the $J^x$.
The scalar and vector squared masses are the eigenvalues of the following matrices:
\begin{align}
M^{(scal)}{}_{I}{}^J&= \left.\frac{1}{2}\frac{\partial^2 V}{\partial\phi^I\partial \phi^K}\,g^{KJ}\right\vert_{\phi^I=0}\,,\nonumber\\
M^{(vec)}{}_{M}{}^{N}&= -\left.2 k_M^u\,h_{uv}\,k_P^v\,\mathcal{M}^{PN}\right\vert_{\phi^I=0}\,,
\end{align}
where by $\phi^I$ we have denoted all 12 scalar fields and where we have written the scalar kinetic term as $g_{IJ}\partial_\mu\phi^I\,\partial^\mu\phi^J$. The matrix $\mathcal{M}_{PN}$ is the symmetric, symplectic, negative-definite matrix characterizing the special K\"{a}hler manifold. Notice that  $M^{(vec)}{}_{M}{}^{N}$ is a $(2n_v)\times (2n_v)$ matrix, where $n_v=3$ is the number of vectors. This is due to the redundant symplectic-covariant notation adopted, so that half the eigenvalues of $M^{(vec)}{}_{M}{}^{N}$ are always 0, while the remaining ones, which depend on the duality frame, being the squared  masses of the vector fields.\par
We further require the gauged model to admit a truncation to the dilatons. This restricts the choice of the generators within the ${\rm G}_{2(2)}$-equivalence classes. A generic element $g\in {\rm G}_{2(2)}$ can be uniquely written as the product of an element $s$ generated by the maximal solvable Borel subalgebra of $\mathfrak{g}_{2(2)}$ and an element $h$ in the maximal compact subgroup (Iwasawa decomposition): $g=s\cdot h$. If a gauging with generators $X_1,\,X_2$ admits a truncation to the dilatons, the gauging of the $g$-transformed generators $g^{-1}\cdot X_1\cdot g,\,g^{-1}\cdot X_2\cdot g$ will in general not have the same property.\par We start by considering some specific gaugings, namely those with gauge generators inside the maximal   $\mathfrak{sl}(2,\mathbb{R})_0\oplus \mathfrak{sl}(2,\mathbb{R})_1$ subalgebra. These will include the model with the $\alpha=1$ Starobinsky potential and one single massive vector multiplet as well as those in which two non-compact Cartan generators are gauged. The latter are the only models, as we show below,  allowing for two massive vector multiplets.  In what follows we shall call
\emph{hyperbolic}, \emph{elliptic} and \emph{parabolic}, $\mathfrak{g}_{2(2)}$ generators whose matrix representation is diagonalizable with real eigenvalues, diagonalizable with imaginary eigenvalues and non-diagonalizable nilpotent, respectively.
 \subsection{Model with \texorpdfstring{$\alpha=1$}{alpha=1} potential and model with two massive vector multiplets }
In the present section we construct the two models mentioned above, namely the one with the $\alpha=1$ potential and the one with two massive vector multiplets. Both can be obtained by gauging isometries in the maximal $\mathfrak{sl}(2,\mathbb{R})_0\oplus \mathfrak{sl}(2,\mathbb{R})_1$ subalgebra of $\mathfrak{g}_{2(2)}$ whose generators can be chosen as follows:
\begin{align}
\mathfrak{sl}(2,\mathbb{R})_0&=\{{\bf T}_0,{\bf T}_+,{\bf T}_-\}\,\,;\,\,\,\,\mathfrak{sl}(2,\mathbb{R})_1=\{{\bf G}_0,{\bf G}_+,{\bf G}_-\}\,,\nonumber\\
{\bf T}_+&= \frac{1}{6}\,e_6\,\,;\,\,\,{\bf T}_-=\frac{1}{6}\,f_6\,\,;\,\,\,{\bf T}_0=2 h_1+h_2\,,\nonumber\\
{\bf G}_+&= e_2\,\,;\,\,\,{\bf G}_-=f_2\,\,;\,\,\,{\bf G}_0=h_2\,,\nonumber\\
[{\bf T}_0,{\bf T}_\pm]&=\pm 2\, {\bf T}_\pm\,\,\,;\,\,\,[{\bf T}_+,{\bf T}_-]={\bf T}_0\,\,;\,\,\,
[{\bf G}_0,{\bf G}_\pm]=\pm 2\, {\bf G}_\pm\,\,\,;\,\,\,[{\bf G}_+,{\bf G}_-]={\bf G}_0\,.
\end{align}
\paragraph{Model with two massive vector multiplets: $X_1$ Hyperbolic, $X_2$ Hyperbolic.}
This case consists in gauging a non-compact Cartan subalgebra of $\mathfrak{g}_{2(2)}$.
We start identifying:
\begin{equation}
X_1=g_1\,({\bf T}_++{\bf T}_-)\,\,;\,\,\,X_2=\frac{g_2}{3}\,({\bf G}_++{\bf G}_-)\,.
\end{equation}
The scalar potential can be truncated to the dilatons $\varphi$ and $U$ and reads:
\begin{equation}
V=\frac{1}{2}\left(g^2_1\,\sinh(2U)^2+g^2_2\,\sinh(\varphi)^2\right)\,.
\end{equation}
The above potential can be truncated to any of the two dilatons.
The scalar Lagrangian density, truncated to the dilatons, reads:
\begin{equation}
e^{-1}\mathcal{L}=\frac{R}{2}-\partial_\mu U\partial^\mu U-\frac{3}{4}\,\partial_\mu \varphi\partial^\mu \varphi-\frac{1}{2}\left(g^2_1\,\sinh(2U)^2+g^2_2\,\sinh(\varphi)^2\right)\,.
\end{equation}
The vacuum is $\mathcal{N}=2$ and the scalar and vector squared masses are (not considering the Goldstones and the massless graviphoton):
\begin{align}
m^2_{scal}\,\,&=\,\,\,5\times (2g_1^2)\,,\,\,5 \times (\frac{2}{3}\,g_2^2)\,,\nonumber\\
m^2_{vecs}\,\,&=\,\,\,2g_1^2\,,\,\,\,\frac{2}{3}\,g_2^2\,.
\end{align}
The scalar $\varphi$ acquires mass-squared $\frac{2}{3}\,g_2^2$, while $U$ the mass-squared $2g_1^2$.
\paragraph{Model with $\alpha=1$ potential.}
We take one of the two generators to be
\begin{equation}
X_1=\frac{g_1}{3}\,\left({\bf G}_+-\frac{3}{2}({\bf T}_+-{\bf T}_-)\right)\,.
\end{equation}
 The only generators commuting with it are ${\bf G}_+$ and ${\bf T}_+-{\bf T}_-$. However, of the two, only the above combination is orthogonal to the $\mathfrak{su}(2)_R$ algebra
thus guaranteeing an $\mathcal{N}=2$ vacuum at the origin. We then find $X_2\propto X_1$, so that only one vector field is effectively gauged. Taking  $X_2$ to be
 \begin{equation}
X_2=\frac{g_2}{3}\,\left({\bf G}_+-\frac{3}{2}({\bf T}_+-{\bf T}_-)\right)\,.
\end{equation}
we find and $\mathcal{N}=2$ vacuum at the origin and a potential, truncated to $\varphi$ alone, of the form:
\begin{equation}
V=\frac{g_1^2+g_2^2}{8}\left(1-\rme^{-\varphi}\right)^2\,.
\end{equation}
The Lagrangian density, truncated to $\varphi$, reads:
\begin{equation}
e^{-1}\mathcal{L}_\varphi=\frac{R}{2}-\frac{3}{4}\,\partial_\mu \varphi\partial^\mu \varphi-\frac{g_1^2+g_2^2}{8}\left(1-\rme^{-\varphi}\right)^2\,.
\end{equation}
On the $\mathcal{N}=2$ vacuum there is one massive vector multiplet of mass squared $m^2=(g_1^2+g_2^2)/6$ which contains the scalaron $\varphi$.
\subsection{A general property of \texorpdfstring{${\rm G}_{2(2)}$}{G2(2)}-gaugings}\label{parabola}
Here we prove that the only gauging  yielding two massive vector fields must involve two semisimple (more specifically hyperbolic) ${\rm G}_{2(2)}$-generators.
As a consequence of this, one cannot find a gauging yielding a Starobinsky-like potential and two massive vector fields. Indeed if $t_\alpha$ is a semisimple generator, the dependence of the  corresponding momentum map on the dilatons $\vec{\phi}=\{U,\varphi\}$ has the following general form:
\begin{equation}
\mathcal{P}_\alpha^x=\sum_I \left(c_{\alpha I}^x\,e^{\vec{a}_I\cdot \vec{\phi}}+d_{\alpha I}^x\,e^{-\vec{a}_I\cdot \vec{\phi}}\right)\,,
\end{equation}
where, for each $I$, $c_{\alpha I}^x\,d_{\alpha I}^x\neq 0$ and  the coefficients $c_{\alpha I}^x,\,d_{\alpha I}^x$ have opposite signs if $t_\alpha$ is hyperbolic, same signs if it is elliptic. Therefore, for each term containing an exponential $e^{\vec{a}_I\cdot \vec{\phi}}$, there is a term containing its inverse $e^{-\vec{a}_I\cdot \vec{\phi}}$. This general property excludes a Starobinsky-like potential.\par
To prove the initial statement we consider a gauging of two commuting ${\rm G}_{2(2)}$-generators of which one is not semisimple and prove that the requirement of an $\mathcal{N}=2$ Minkowski vacuum always implies the two generators to be proportional, thus yielding only one massive vector field.\par
Let us consider the gauging of a non-semisimple quaternionic isometry  generator, represented by a non-diagonalizable matrix.
A non-diagonalizable matrix $X$, by Jordan decomposition, can be written as the sum two commuting matrices: a semisimple (i.e. represented by a diagonalizable matrix) $X^{(0)}$ and a non-vanishing nilpotent one $X^{(N)}$:
\begin{equation}
X=X^{(0)}+X^{(N)}\,\,\,,\,\,\,\,[X^{(0)},X^{(N)}]=0\,.
\end{equation}
Nilpotent $\mathfrak{g}_{2(2)}$ generators fall into 5 orbits  under the adjoint action of ${\rm G}_{2(2)}$ (see for instance \cite{Collingwood:1993,Kim:2010bf,Fre:2011uy}):
\begin{enumerate}
\item{$\mathcal{O}_1$ whose representative is $e_6$. The little group is $G_\ell={\rm SL}(2,\mathbb{R})\ltimes \mathbb{R}^5$ and the orbit is 6-dimensional. In the ${\bf 7}$ representation of $\mathfrak{g}_{2(2)}$, its elements are nilpotent of degree 2;}
    \item{$\mathcal{O}_2$ whose representative is $e_2$. The little group is $G_\ell={\rm SL}(2,\mathbb{R})\ltimes \mathbb{R}^3$ and the orbit is 8-dimensional. In the ${\bf 7}$ representation of $\mathfrak{g}_{2(2)}$, its elements are nilpotent of degree 2;}
\item{$\mathcal{O}_3$ whose representative is $e_2+e_6$. The little group is $G_\ell=\mathbb{R}^4$ (non-abelian) and the orbit is 10-dimensional. In the ${\bf 7}$ representation of $\mathfrak{g}_{2(2)}$, its elements are nilpotent of degree 3;}
    \item{$\mathcal{O}_4$ whose representative is $e_2-e_6$. The little group is $G_\ell=\mathbb{R}^4$ (non-abelian) and the orbit is 10-dimensional. In the ${\bf 7}$ representation of $\mathfrak{g}_{2(2)}$, its elements are nilpotent of degree 3;}
   \item{$\mathcal{O}_5$ whose representative is $e_1+e_2$. The little group is $G_\ell=\mathbb{R}^2$ (abelian) and the orbit is 12-dimensional. In the ${\bf 7}$ representation of $\mathfrak{g}_{2(2)}$, its elements are nilpotent of degree 7;}
\end{enumerate}
In what follows we choose the nilpotent component $X^{(N)}_1$ of the first generator $X_1$ in each of the above orbits and choose the remaining semisimple component in the algebra of the corresponding little group. The second  generator $X_2$ is  chosen to commute with the first. Both generators are then constrained by requiring the existence of an $\mathcal{N}=2$ vacuum.
\subsubsection{Orbit \texorpdfstring{$\mathcal{O}_1$}{O1}}
As a representative of the orbit we can choose $X_1^{(N)}\propto {\bf T }_+=e_6/6$.
The little algebra $\mathfrak{g}_\ell$ of $G_\ell$ is the semi-direct sum of an $\mathfrak{sl}(2,\mathbb{R})_\ell$ part and an abelian $\mathbb{R}^3$:
\begin{equation}
\mathfrak{sl}(2,\mathbb{R})_\ell=\mathfrak{sl}(2,\mathbb{R})_1\,\,;\,\,\,\mathbb{R}^5={\rm Span}(e_1,\,e_3,\,e_4,\,e_5,\,e_6)\,.
\end{equation}
We have verified by direct computation, using a MATHEMATICA code,  that all representatives of $\mathcal{O}_1$ have non-vanishing momentum maps at the origin. This is done by computing first $s^{-1}\cdot X_1^{(N)}\cdot s$, with $s$ in the solvable group generated by the Borel subalgebra, and verifying that its momentum map is always non-vanishing for any $s$. This property clearly does not change if we further conjugate the representative by a compact transformation. To have a generator $X_1$ with a vanishing momentum map  at the origin, we therefore need to combine $X_1^{(N)}$ with a semisimple generator  $X_1^{(0)}$ in $\mathfrak{g}_\ell$, namely in the algebra $\mathfrak{sl}(2,\mathbb{R})_1$, and  we choose $X_2$ so that $[X_1,\,X_2]=0$. We find that $X_2$ can only be a combination of $X_1^{(N)}$ and $X_1^{(0)}$. The existence of an $\mathcal{N}=2$ Minkowski vacuum further restricts this combination in both $X_1$ and $X_2$ so that
\begin{equation}
X_1\propto X_2\,,\label{conclusion1}
\end{equation}
which implies the existence of a single massive vector field on the vacuum.  This conclusion, in the light of the above discussion, clearly would not change had we started from a different representative $X^{(N)}$ of the orbit.\par
%We can consider the effect on this result of a conjugation by an element $g$ of ${\rm G}_{2(2)}$. As explained in  Sect. \ref{sect1}, we can Iwasawa-decompose $g$ as $g=s\cdot h$, $s$ solvable and $h$ compact. A conjugation by $h$ will not alter the above conditions (in particular the susy one which involves the trace with the $\mathfrak{su}(2)_R$ generators), yielding the same conclusion. As for the action by $s$, we have:
%\begin{equation}
%s^{-1}\cdot e_6\cdot s\propto e_6\,,
%\end{equation}
%so that it will not affect the above conclusions (\ref{conclusion1}).\par
A specific example of this gauging corresponds to choosing:
\begin{equation}
X_1=g_1\,\left({\bf T}_+-\frac{{\bf G}_+-{\bf G}_-}{6}\right)\,\,;\,\,\,X_2=g_2\,\left({\bf T}_+-\frac{{\bf G}_+-{\bf G}_-}{6}\right)\,.
\end{equation}
The model admits a consistent truncation to the dilatons and, in particular, to $U$ alone, yielding:
\begin{equation}
V=\frac{1}{8}\,(g_1^2+g_2^2)\,(1-\rme^{-2U})^2\,.\label{potU}
\end{equation}
This potential corresponds to the class of the $\alpha$ attractor models (see Subsection \ref{cp1uni} below) with $\alpha=1/3$. In the resulting $\mathcal{N}=2$ vacuum at the origin, we have one massive vector multiplet with mass-squared $(g_1^2+g_2^2)/2$, and one massless vector multiplet, with a massless hypermultiplet.
\subsubsection{Orbit \texorpdfstring{$\mathcal{O}_2$}{O2}}
As a representative of the orbit we start choosing $X_1^{(N)}\propto {\bf G }_+=e_2$. The little algebra $\mathfrak{g}_\ell$ of $G_\ell$ is the semi-direct sum of an $\mathfrak{sl}(2,\mathbb{R})_\ell$ part and an abelian $\mathbb{R}^3$:
\begin{equation}
\mathfrak{sl}(2,\mathbb{R})_\ell=\mathfrak{sl}(2,\mathbb{R})_0={\rm Span}(2h_1+h_2,\,e_6,\,f_6)\,\,;\,\,\,\mathbb{R}^3={\rm Span}(e_2,\,e_5,\,f_1)\,.
\end{equation}
The semisimple component $X_1^{(0)}$ of $X_1$ will then belong to $\mathfrak{sl}(2,\mathbb{R})_0$.
Next we choose $X_2$ so that $[X_1,\,X_2]=0$. We find that $X_2$ can only be a combination of $X_1^{(N)}$ and $X_1^{(0)}$. Moreover the momentum map of any representative of $\mathcal{O}_2$ is never zero, as it can be verified by computing the momentum map at the origin of $s^{-1}\, e_2\,s$, $s$ being a generic element of the Borel subgroup. This clearly does not change if we further transform the generator by any compact transformation. Therefore the requirement of  $\mathcal{N}=2$ supersymmetry at the origin, for any choice of $X_1^{(N)}\in \mathcal{O}_2$ always requires $X_1$ to be a well defined combination of its semisimple and nilpotent parts. $X_2$ will then be proportional to the same combination.
The existence of an $\mathcal{N}=2$ Minkowski vacuum therefore requires  both $X_1$ and $X_2$ to be proportional
\begin{equation}
X_1\propto X_2\,,\label{conclusion2}
\end{equation}
which excludes two massive vector multiplets.
An example of such gauging is the one yielding a Starobinsky potential, though with only one massive vector multiplet,  discussed earlier.\par
%Just as in the previous case, we now analyze  the effect of a ${\rm G}_{2(2)}$-conjugation on the gauge generators. It can be verified that no element of $\mathcal{O}_2$ can alone yield an $\mathcal{N}=2$ vacuum.  As in the example discussed above, it has to be combined by an element of the corresponding little algebra. Conjugation of the two gauge generators  by a compact element will not alter the conditions yielding (\ref{conclusion2}).
%We therefore consider the effect of a solvable element $s$ in $\exp(Borel)$.  The effect of a solvable transformation $s$, however, will in general spoil truncability to the dilatons.
\subsubsection{Orbit \texorpdfstring{$\mathcal{O}_3$}{O3}}
We take as representative $X_1^{(N)}\propto e_2+e_6$. The little algebra is
\begin{equation}
\mathfrak{g}_\ell=\mathbb{R}^4={\rm Span}(e_2+e_6,\,e_2-e_6,\,e_4-f_1,\,e_5)\,,
\end{equation}
which is non-abelian and   $e_2+e_6$ is a central charge. Since there is no semisimple generator commuting with $X_1^{(N)}$, $X_1^{(0)}=0$. Moreover this orbit is incompatible with the existence of an $\mathcal{N}=2$ Minkowski vacuum. We have indeed checked that the momentum map at the origin of the transformed $s^{-1}\cdot X_1\cdot s$ generator  under a generic transformation $s$ in the Borel subgroup is never zero. This clearly does not change if we further transform the generator by any compact transformation.
Therefore no representative of the ${\rm G}_{2(2)}$-orbit of $X_1$ admits an $\mathcal{N}=2$ Minkowski vacuum.
\subsubsection{Orbit \texorpdfstring{$\mathcal{O}_4$}{O4}}
We take as representative $X_1^{(N)}\propto e_2-e_6$ whose little algebra is
\begin{equation}
\mathfrak{g}_\ell=\mathbb{R}^4={\rm Span}(e_2+e_6,\,e_2-e_6,\,e_4+f_1,\,e_5)\,,
\end{equation}
in which $e_2-e_6$ is the central charge. Just as for the previous orbit, since there is no semisimple generator commuting with $X_1^{(N)}$, $X_1^{(0)}=0$.  This time, however, there are representatives  of the orbit that are compatible with the existence of an  $\mathcal{N}=2$ Minkowski vacuum (in fact there is a two-parameter family of such representatives). For each of these representatives $X_1$, the most general $X_2$ commuting with it and preserving $\mathcal{N}=2$ supersymmetry must be proportional to $X_1$ itself:
\begin{equation}
X_1\propto X_2\,.\label{conclusion3}
\end{equation}
We can take, for instance,
\begin{equation}
X_1=g_1(e_6-2 e_2)\in \mathcal{O}_4\,\,;\,\,\,\,X_2=g_2\,(e_6-2 e_2)\in \mathcal{O}_4\,,
\end{equation}
yielding, upon truncation to the dilatons, the scalar potential:
\begin{equation}
V=\frac{9}{2} \rme^{-2 (2 U+\varphi )} \left(\rme^{2
   U}-\rme^{\varphi }\right)^2
   \left(g_1^2+g_2^2\right)\,,
\end{equation}
which has an extremum  at the origin with  $\mathcal{N}=2$ susy and  a single massive vector multiplet with squared mass $m^2=24\,(g_1^2+g_2^2)$.
\subsubsection{Orbit \texorpdfstring{$\mathcal{O}_5$}{O5}}
We choose as representative $X_1^{(N)}=e_1+e_2$, whose little algebra is:
\begin{equation}
\mathfrak{g}_\ell=\mathbb{R}^2={\rm Span}(e_1+e_2,\,e_6)\,,
\end{equation}
and is abelian. Also in this case $X_1=X_1^{(N)}$. Moreover there is a three-parameter family of representatives $X_1$ of $\mathcal{O}_5$ that are compatible with the existence of an $\mathcal{N}=2$ Minkowski vacuum. Just as for the previous orbit, the supersymmetry condition on $X_2\in \mathfrak{g}_\ell$ restricts it to be proportional to $X_1$, thus yielding a single massive vector multiplet.
\subsection{Models with  one vector-multiplet and one hypermultiplet}
Now let us consider the case in which one massive vector multiplet (the one not containing the scalaron)
has been integrated out (infinite mass limit) so that we are left with a $\mathcal{N}=2$ supergravity coupled to one (massless) vector multiplet and one (massless) hypermultiplet. We wish now to generate a massive vector multiplet on an $\mathcal{N}=2$ Minkowski vacuum through the  gauging of an isometry in the quaternionic-K\"{a}hler (QK) manifold. The scalar manifold of this minimal model has the form:
\begin{align}
\mathcal{M}_{scal}&=\mathcal{M}_{SK}\times \mathcal{M}_{QK} \,,\nonumber\\
\mathcal{M}_{SK}&=\frac{{\rm SL}(2,\mathbb{R})}{{\rm SO(2)}}\,\,;\,\,\,\mathcal{M}_{QK}=\begin{cases}\frac{{\rm SU}(2,1)}{{\rm U(2)}}\cr \frac{{\rm SO}(1,4)}{{\rm SO(4)}}\end{cases}\,,
\end{align}
where the special K\"{a}hler (SK) manifold can either be the cubic one of the $T^3$ model, or $\mathbb{CP}^1$.
In none of the four different combinations we find a consistent truncation to a single Cartan scalar yielding the Starobinsky model.
However, only if $\mathcal{M}_{SK}$ is  $\mathbb{CP}^1$, we can consistently truncate to the hyperscalars, and find ``Starobinsky-like'' models.\par
%As for $\mathcal{M}_{SK}$ we choose the cubic manifold of the $T^3$ model since the dilatonic scalar  the QK manifolds does not have the correct normalization of the kinetic term in order to be identified with the scalaron. The dilaton in the imaginary part of the complex scalar $T=y-i \rme^\varphi$ has, on the other hand, the correct normalization of the kinetic term.\par
As we gauge a single  isometry  $g$ of the QK manifold, the gauge generators $X_M$ will have the simple form: $X_M=\theta_M\,g$, so that, if we denote by $k^u$ and $\mathcal{P}^x$ the  Killing vector  and momentum map associated with $g$, respectively, we can write:
\begin{equation}
k_M^u=\theta_M\,k^u\,\,\,;\,\,\,\,\mathcal{P}^x_M=\theta_M\,\mathcal{P}^x\,.
\end{equation}
The embedding tensor, in other words, can be thought of as a single vector of ``electric-magnetic'' charges $\theta_M$.
To rewrite the  scalar potential it is useful to define ``central charges'', $Z,\,Z_i$ (in our case $i$ has a single value) defined as follows:
\begin{equation}
Z=V^M\,\theta_M\,\,;\,\,\,Z_i=D_iV^M\theta_M\,.
\end{equation}
The scalar potential reads:
\begin{equation}
V=(4\,k^2-3 \mathcal{P}^2)\,|Z|^2+\mathcal{P}^2\,Z_i g^{i\bar {\imath}}Z_{\bar {\imath}}\,,
\end{equation}
where $k^2\equiv k^u\,h_{uv}\,k^v$, $\mathcal{P}^2\equiv \mathcal{P}^x \mathcal{P}^x$.\par
The existence of a Minkowski $\mathcal{N}=2$ vacuum at the origin, with ungauged central charge, implies $\mathcal{P}^x=0$ and $V^M(0)\,\theta_M=0$.

\subsection{\texorpdfstring{$\mathbb{CP}^1\times $}{CP1 x} universal}\label{cp1uni}
As for $\mathcal{M}_{SK}$ we choose $\mathbb{CP}^1$ which allows for a truncation to the QK scalars.
In this case the most general gauged Lagrangian yielding an $\mathcal{N}=2$ Minkowski vacuum, upon truncation to the QK dilaton $U$, reads:
\begin{equation}
e^{-1}\,\mathcal{L}=\frac{R}{2}-\partial_\mu U \partial^\mu U-V(U)\,\,;\,\,\,V(U)=g^2\,\rme^{-4 U}(1-\rme^{2U})^2\left[7 c_1+ 8 c_2 +\rme^{2U}\, (c_1-8 c_2)\right]^2\,,
\end{equation}
where $c_i$ are constants related to the choice of the gauge generator.
If $c_1=8 c_2$, the gauge generator is non-semisimple (i.e. parabolic) and  the potential is ``Starobinsky-like'':
\begin{equation}
e^{-1}\,\mathcal{L}=\frac{R}{2}-\partial_\mu U \partial^\mu U-V(U)\,\,;\,\,\,V(U)=g^2\,(1-\rme^{-2U})^2\,,\label{abovepot}
\end{equation}
which has the same form as the potential (\ref{potU}) found in a specific $G_{2(2)}$-gauging.
To compare with the literature about the $\alpha$-attractors, we should reduce the above Lagrangian density to the canonical asymptotic form for $\phi\rightarrow \infty$:
\begin{equation}
e^{-1}\,\mathcal{L}=\frac{R}{2}-\frac{1}{2}\partial_\mu \phi \partial^\mu \phi-V_0\,\left(1-\rme^{-\sqrt{\frac{2}{3\alpha}}\phi}+\dots\right)\,.\label{alphat}
\end{equation}
The potential (\ref{abovepot}) corresponds then to the value $\alpha=1/3$.
\subsection{\texorpdfstring{$\mathbb{CP}^1\times \mathbb{HP}^1$}{CP1 x HP1}}
In this case the most general gauged Lagrangian yielding with an $\mathcal{N}=2$ Minkowski vacuum, upon truncation to the QK dilaton $U$, reads:
\begin{equation}
e^{-1}\,\mathcal{L}=\frac{R}{2}-\frac{1}{2}\,\partial_\mu U \partial^\mu U-V(U)\,\,;\,\,\,V(U)=g^2\,\rme^{-2 U}(1-\rme^{U})^2\left[-2 c_1+ 3 c_2 +\rme^{U}\, (2c_1+ c_2)\right]^2\,.
\end{equation}
If $c_1=- c_2/2$, the gauge generator is non-semisimple and the potential is ``Starobinsky-like'' and we find:
\begin{equation}
e^{-1}\,\mathcal{L}=\frac{R}{2}-\frac{1}{2}\,\partial_\mu U \partial^\mu U-V(U)\,\,;\,\,\,V(U)=g^2\,(1-\rme^{-U})^2\,,
\end{equation}
yielding an $\alpha$-attractor with $\alpha=2/3$.
\section{Conclusions}
In this work we have started a systematic inspection of gauged $\mathcal{N}=2$ supergravity potentials that are suitable to describe inflationary regimes. The work was originally inspired by the  Planck data which seem to favor Starobinsky-like potentials and the problem of embedding these scenarios in extended supergravities, which have more direct links to superstring theories.
 Extended supersymmetry on the other hand puts more stringent constraints on the field content of the model and its interactions.
However, our general setup and  results will  have a bearing also on different (e.g. non-minimal) constructions  which can be inspired by more recent results, like the BICEP2 ones, or other future cosmological observations.
\section*{Acknowledgements}
This work was supported in part by the ERC Advanced Grants no.226455 (SUPERFIELDS), by the MIUR grant RBFR10QS5J, by the FWO - Vlaanderen, Project No. G.0651.11, by the Interuniversity Attraction Poles Programme initiated by the Belgian Science Policy (P7/37), and by the COST Action MP1210 "The String Theory Universe".
\appendix
\section{Conformal dualization of \texorpdfstring{$R+R^2$}{R+R2} action}
\label{app:confdual}
The conformal dualization can be seen as follows. We start from the conformal-invariant action
\begin{equation}
S=\int \rmd^4 x\,\sqrt{g}  \left[   -\frac12\phi\bbox^C \phi +36\alpha  \left(\frac{\bbox^C \phi}{\phi }\right)^2\right]\,, \label{Sconformalalpha}
\end{equation}
where $\bbox^C \phi$ is the conformal d'Alembertian  %XXX here should still come an original ref. XXX,
(see e.g. \cite{Freedman:2012zz}). We will explain that it is equivalent to the $R+R^2$ action, and that by changing the dilatational gauge, it can be written as a model with a scalar and the Starobinsky potential. In a dilatational gauge
\begin{equation}
  \phi =\frac{\sqrt{6}}{\kappa }\,\rightarrow \, \bbox^C \phi=-\frac{1}{\sqrt{6}\kappa }R\,,\qquad \frac{\bbox^C \phi}{\phi }=-\frac{1}{6}R\,,
 \label{phigaugePoinc}
\end{equation}
the action (\ref{Sconformalalpha}) reduces to (\ref{SR2bos}) (without the $\beta $-term). We can rewrite (\ref{Sconformalalpha}) using an auxiliary field $\chi $ and a Lagrange multiplier field $\sigma $ as
\begin{equation}
S=\int \rmd^4 x\,\sqrt{g}  \left[ -\frac12\phi\bbox^C \phi+\sigma \left(\chi -\frac{\bbox^C \phi}{\phi }\right)+36\alpha\,\chi ^2\right] \,.
\label{chiLagrangef}
\end{equation}
After elimination of $\chi $, this can be written as
\begin{equation}
S=\int \rmd^4 x\,\sqrt{g}  \left[-(\frac12\phi ^2+\sigma)\frac{\bbox^C \phi}{\phi }-\frac{1}{144\,\alpha }\sigma ^2\right] \,.
\label{elimchiLagrangef}
\end{equation}
Using now a dilatation gauge
\begin{equation}
  \frac12\phi ^2+\sigma=\frac{3}{\kappa ^2}\,,
 \label{dilgauge2}
\end{equation}
this action reduces to
\begin{equation}
  S=\int \rmd^4 x\,\sqrt{g}  \left[ \frac{1}{2\kappa ^2}R -\frac{3(\partial _\mu \phi)(\partial ^\mu \phi) }{(\kappa \phi) ^2}-\frac{1}{16\,\alpha\kappa ^4 }\left(1-\ft16\kappa ^2\phi ^2\right) ^2\right] \,,
 \label{Swithmassivephi}
\end{equation}
which can be canonically normalized with $\kappa \phi = \sqrt{6}\exp\left( -\frac{1}{\sqrt{6}}\kappa \varphi \right)$.
Then the action takes the form
\begin{equation}
  S=\int \rmd^4 x\,\sqrt{g}  \left[ \frac{1}{2\kappa ^2}R -\frac12(\partial _\mu \varphi )(\partial ^\mu \varphi)-\frac{1}{16\,\alpha\kappa ^4 }\left[1-\exp\left( -\sqrt{\frac{2}{3}}\kappa \varphi \right)\right]^2\right]  \,.
  \label{Starobinskyscalar}
\end{equation}
It shows that the compensating scalar has become a massive scalar field with mass $m_0$, such that $m_0{}^{-2}=12\alpha \kappa ^2$.

\section{\texorpdfstring{$\mathfrak{g}_{2(2)}$}{g2(2)} generators}\label{g2}
\begin{figure}[ht]
\begin{center}
\centerline{\includegraphics[width=0.8\textwidth]{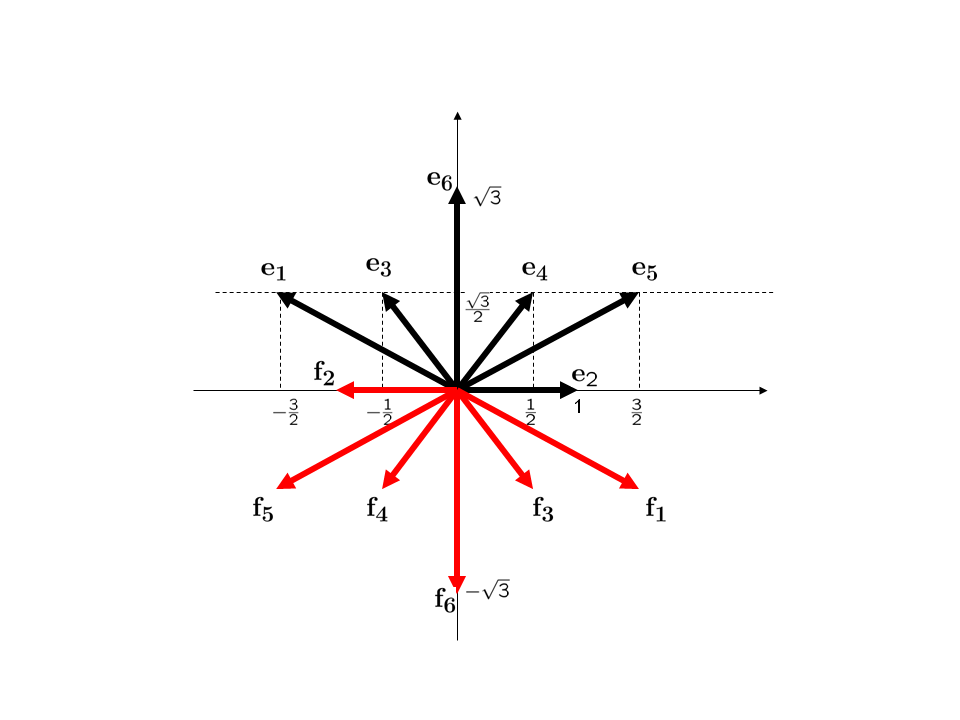}}
\caption{}\label{fig1}
\end{center}
\end{figure}

Let us describe the  $\mathfrak{g}_{2(2)}$ generators in the Chevalley basis: $h_1,h_2,e_i,f_i$, $i=1,\dots, 6$. The roots corresponding to the nilpotent generators $e_i,\,f_i$ are drawn in Fig. \ref{fig1} and satisfy the relations:
\begin{align}
[h_1,e_1]&=2e_1\,\,;\,\,\,[h_2,e_2]=2e_2\,\,;\,\,\,[h_1,e_2]=-e_2\,\,;\,\,\,[h_2,e_1]=-3e_1\,\,;\nonumber\\ [e_2,f_2]&=h_2\,\,;\,\,\,[e_1,f_1]=h_1\,\,;\,\,\,
[e_1,f_2]=0\,\,;\,\,\,[e_2,f_1]=0\,\,;\nonumber\\
[e_1,e_2]&=e_3\,\,;\,\,\,[e_4,e_2]=e_5\,\,;\,\,\,[e_1,e_5]=e_6\,,
\end{align}
the remaining relations can be inferred from the action of the Cartan involution $\tau$ (which in the real ${\bf 7}$ representation we work with, amounts to $\tau(X)=-X^T$): $\tau(h_i)=-h_i$, $\tau(e_i)=-f_i$.
The parametrization of the quaternionic manifold is defined by the following coset representative:
\begin{align}
\mathbb{L}&=\exp\left(- \frac{\chi}{6}\, \,e_6\right)\,\exp\left(\sqrt{2}  \zeta^0\,{e_1}-\sqrt{2}  \zeta^1\,{e_3}+\frac{
   \hat{\zeta }_0}{3 \sqrt{2}}\,{e_5}+\frac{\hat{\zeta }_1}{3 \sqrt{2}}\,{e_4} \right)\,\exp\left(y\,e_2\right)\,\exp\left(\frac{1}{2}\varphi\,{\bf G}_0 \right)\times\nonumber\\
   &\times \exp\left(U\,{\bf T}_0\right)\,.
\end{align}
The two dilatons are $\varphi,\,U$. The former defines the imaginary part of the complex coordinate $T=y-\rmi\,\rme^{\varphi}$ spanning a cubic $T^3$-submanifold of the quaternionic one.\par
The $\mathfrak{su}(2)$ R--symmetry subalgebra generators $J^x$ read:
\begin{align}
J^1&=-\frac{1}{12}(6 {e_3}- {e_5}-6  {f_3}+ {f_5})\,\,,\,\,\,J^2=-\frac{1}{12}(6 e_1-3 e_4-6 f_1+3 f_4)\,,\nonumber\\J^3&
=-\frac{1}{12}(-6 e_2-e_6+6 f_2+f_6)\,,\nonumber\\
[J^x,J^y]&= 2\,\epsilon^{xyz}\,J^z\,.
\end{align}
We also have an $\mathfrak{su}(2)$ subalgebra commuting with the  above R--symmetry one, whose generators $\tilde{J}^x$ have the following form:
\begin{align}
\tilde{J}^1&=\frac{1}{4} (6  {e_1}+ {e_4}-6  {f_1}- {f_4})\,\,;\,\,\tilde{J}^2=\frac{1}{4} (-2  {e_3}- {e_5}+2 {f_3}+ {f_5})\,\,;\,\,\tilde{J}^3=\frac{1}{4} (2  {e_2}- {e_6}-2  {f_2}+ {f_6})\,.
\end{align}
The momentum maps are given by:
\begin{equation}
\mathcal{P}_\alpha{}^x=\frac{1}{2}\,{\rm Tr}\left(J^x\,\mathbb{L}^{-1}\,t_\alpha\,\mathbb{L}\right)\,,
\end{equation}
where all generators are computed in the ${\bf 7}$ of ${\rm G}_{2(2)}$.\par
It is useful to give the inner product of the generators in the chosen basis, defined in terms of the trace. The non vanishing
products are:
\begin{align}
{\rm Tr}(h_1\,h_1)&=4\,\,;\,\,\,{\rm Tr}(h_1\,h_2)=-6\,\,;\,\,\,{\rm Tr}(h_2\,h_2)=12\,\,;\\ {\rm Tr}(e_1\,f_1)&=2\,\,;\,\,\,{\rm Tr}(e_2\,f_2)={\rm Tr}(e_3\,f_3)=6\,;\,\,\,
{\rm Tr}(e_4\,f_4)&=24\,\,;\,\,\,{\rm Tr}(e_5\,f_5)={\rm Tr}(e_6\,f_6)=72\,.\nonumber
\end{align}
Below we give the Killing vectors and  momentum maps restricted to the Cartan fields:
{\small \begin{align}
K_{J^1}&=\frac{\rme^{-\varphi }}{2 \sqrt{2}}\, \left(\left(-\rme^{2 U}+\rme^{\varphi }\right) \partial _{\zeta _1}-\rme^{\varphi } \left(-1+\rme^{2 U+3 \varphi
   }\right) \partial _{\hat{\zeta }_0}\right)\,,\nonumber\\
K_{J^2}&=\frac{\rme^{-3 \varphi } }{2 \sqrt{2}}\left(\left(\rme^{2 U}-\rme^{3 \varphi }\right) \partial _{\zeta _0}-3 \rme^{3 \varphi } \left(-1+\rme^{2
   U+\varphi }\right) \partial _{\hat{\zeta }_1}\right)\,,\nonumber\\
K_{J^3}&=\frac{1}{2} \left(\left(-1+\rme^{4 U}\right) \partial _\chi-\left(-1+\rme^{2 \varphi }\right) \partial _y\right)\,,\nonumber\\
K_{\tilde{J}^1}&=\frac{\rme^{-3 \varphi } }{2 \sqrt{2}}\left(-3 \left(\rme^{2 U}-\rme^{3 \varphi }\right) \partial _{\zeta _0}-3 \rme^{3 \varphi } \left(-1+\rme^{2
   U+\varphi }\right) \partial _{\hat{\zeta }_1}\right)\,,\nonumber\\
K_{\tilde{J}^2}&=\frac{\rme^{-\varphi } }{2 \sqrt{2}}\left(\left(-\rme^{2 U}+\rme^{\varphi }\right) \partial _{\zeta _1}+3 \rme^{\varphi } \left(-1+\rme^{2 U+3
   \varphi }\right) \partial _{\hat{\zeta }_0}\right)
   \,,\nonumber\\
K_{\tilde{J}^3}&=\frac{1}{2} \left(-3 \left(-1+\rme^{4 U}\right) \partial _\chi-\left(-1+\rme^{2 \varphi }\right) \partial _y\right) \,,\nonumber\\
K_{h_1}&=\frac{1}{2} \left(\partial _U-2 \rme^{\varphi } \partial _b\right)\,,\nonumber\\
K_{h_2}&=2 \rme^{\varphi } \partial _b\,,\nonumber\\
K_{e_6}&=-6 \partial _\chi\,\,;\,\,\,K_{e_2}=\partial _y\,\,;\,\,\,K_{e_1}=\frac{\partial _{\zeta _0}}{\sqrt{2}}\,\,;\,\,\,K_{e_3}=-\frac{\partial _{\zeta _1}}{\sqrt{2}}\,,\nonumber\\
K_{e_5}&=3 \sqrt{2} \partial _{\hat{\zeta }_0}\,\,;\,\,\,K_{e_4}=3 \sqrt{2} \partial _{\hat{\zeta }_1}\,,\nonumber\\
\mathcal{P}_{J^1}^x&=-\frac{1}{2}\left( 3 \cosh \left(U-\frac{\varphi }{2}\right)+\cosh \left(U+\frac{3 \varphi
   }{2}\right),0,0\right)\,,\nonumber\\
\mathcal{P}_{J^2}^x&=-\frac{1}{2} \left(0,\cosh \left(U-\frac{3 \varphi }{2}\right)+3 \cosh \left(U+\frac{\varphi
   }{2}\right),0\right)\,,\nonumber\\
\mathcal{P}_{J^3}^x&=\left(0,0,\frac{1}{2} (-\cosh (2 U)-3 \cosh (\varphi ))\right)\,,\nonumber\\
\mathcal{P}_{\tilde{J}^1}^x&=\left(0,-3 \sinh \left(U-\frac{\varphi }{2}\right) \sinh (\varphi ),0\right)\,,\nonumber\\
\mathcal{P}_{\tilde{J}^2}^x&=\left(3 \sinh \left(U+\frac{\varphi }{2}\right) \sinh (\varphi ),0,0\right)\,,\nonumber\\
\mathcal{P}_{\tilde{J}^3}^x&=\left(0,0,\frac{3}{2} (\cosh (2 U)-\cosh (\varphi ))\right)\,,\nonumber\\
\mathcal{P}_{h_1}^x&=\mathcal{P}_{h_2}^x=\left(0,0,0\right)\,,\nonumber\\
\mathcal{P}_{e_6}^x&=\left(0,0,-3 \rme^{-2 U}\right)\,\,;\,\,\,\mathcal{P}_{e_2}^x=\left(0,0,-\frac{3 \rme^{-\varphi }}{2}\right)\,\,;\,\,\,\mathcal{P}_{e_1}^x=\left(0,\frac{1}{2} \rme^{\frac{3 \varphi }{2}-U},0\right)\,,\nonumber\\
\mathcal{P}_{e_3}^x&=\left(\frac{3}{2} \rme^{\frac{\varphi }{2}-U},0,0\right)\,\,;\,\,\,\mathcal{P}_{e_5}^x=\left(-3 \rme^{-U-\frac{3 \varphi }{2}},0,0\right)\,\,;\,\,\,\mathcal{P}_{e_4}^x=\left(0,-3 \rme^{-U-\frac{\varphi }{2}},0\right)\,,
\end{align}}
where $b=\rme^{\varphi}$.
\section{\texorpdfstring{$T^3$}{T3} model}\label{T3}
The K\"{a}hler potential and metric has the form:
\begin{equation}
\mathcal{K}=-3\log(\rmi(T-\bar{T}))=-\log(8\,\rme^{3\varphi})\,\,;\,\,\,
g_{T\bar{T}}=-\frac{3}{(T-\bar{T})^2}=\frac{3}{4}\,\rme^{-2\varphi}\,,
\end{equation}
where $T^y-\rmi\,\rme^{\varphi}$.
The symplectic, covariantly holomorphic section $V^M$ reads:
\begin{equation}
V^M=\rme^{\frac{\mathcal{K}}{2}}\,(1,\,T,\,-T^3,\,3\,T^2)\,\,,\,\,D_T V^M=\frac{\rme^{\frac{\mathcal{K}}{2}}}{T-\bar{T}}\,(-3,\,-(2T+\bar{T}),\,3T^2 \bar{T},\,-3\,T(T+2 \bar{T}))\,.
\end{equation}
\section{The Universal Model}\label{univ}
Let us write the generators of the $\mathfrak{su}(2,1)$ as follows:
\begin{equation}
\mathfrak{su}(2,1)={\rm Span}(J^x,\,J_0,\, H_0,\,T_a,\,T_\bullet)\,\,,\,\,\,a=1,2\,,
\end{equation}
where $J^x$ are the quaternionic structure generators, $J_0$ is the ${\rm U}(1)$ generator commuting with them. The four  remaining generators  $ H_0,\,T_a,\,T_\bullet$ generate the Borel subalgebra. The matrix representation of the generators in the fundamental of ${\rm SU}(1,2)$ is:
\begin{align}
J^1&=\left(
\begin{array}{lll}
 0 & 0 & 0 \\
 0 & 0 & -i \\
 0 & -i & 0
\end{array}
\right)\,\,;\,\,\,J^2=\left(
\begin{array}{lll}
 0 & 0 & 0 \\
 0 & 0 & -1 \\
 0 & 1 & 0
\end{array}
\right)\,\,;\,\,\,J^3=\left(
\begin{array}{lll}
 0 & 0 & 0 \\
 0 & -i & 0 \\
 0 & 0 & i
\end{array}
\right)\,\,;\nonumber\\
J_0&=\left(
\begin{array}{lll}
 -\frac{4 i}{3} & 0 & 0 \\
 0 & \frac{2 i}{3} & 0 \\
 0 & 0 & \frac{2 i}{3}
\end{array}
\right)\,\,;\,\,\,
H_0=\left(
\begin{array}{lll}
 0 & 0 & \frac{1}{2} \\
 0 & 0 & 0 \\
 \frac{1}{2} & 0 & 0
\end{array}
\right)\,\,;\nonumber\\
T_1&=\frac{1}{2\sqrt{2}}\,\left(
\begin{array}{ccc}
 0 & -1-i & 0 \\
 -1+i & 0 & 1-i \\
 0 & -1-i & 0
\end{array}
\right)\,\,;\,\,\,T_2=\left(
\begin{array}{ccc}
 0 & 1-i & 0 \\
 1+i & 0 & -1-i \\
 0 & 1-i & 0
\end{array}
\right)\,,\nonumber\\
T_\bullet &=-\frac{i}{2}\,\left(
\begin{array}{lll}
 1 & 0 & -1 \\
 0 & 0 & 0 \\
 1 & 0 & -1
\end{array}
\right)\,,
\end{align}
the invariant matrix defining the fundamental representation is ${\rm diag}(+1,-1,-1)$. The commutation relations among the generators of the Borel subalgebra are:
\begin{equation}
[H_0,T_\bullet]= T_\bullet\,\,,\,\,\,[H_0,T_M]= \frac{1}{2}\,T_M\,\,\,;\,\,\,\,[T_1,\,T_2]=T_\bullet\,.
\end{equation}
The parametrization is defined by a coset representative of the form:
\begin{equation}
\mathbb{L}(q^u)=\rme^{-\chi T_\bullet}\,\rme^{\sqrt{2}Z^M\,T_M}\,\rme^{2 U H_0}\,.
\end{equation}
The metric in the normalization in which $\Omega^x_{uv}=-K^x_{uv}$, corresponding to the normalization of the kinetic term in (\ref{lagra}), reads:
\begin{equation}
ds^2=dU^2+\frac{\rme^{-4U}}{4}(d\chi+Z^T\mathbb{C}dZ)^2+\frac{\rme^{-2U}}{2}dZ^T\,dZ\,.
\end{equation}
The corresponding scalar curvature is $\mathcal{R}=-24=-8n(n+2)$ ($n=1$).\par
The Killing vectors are:
\begin{align}
K_{J^1}&=\frac{(Z^1+Z^2)}{2}\partial_U+\frac{1}{2}\,\left[{\rm Im}(\mathcal{E}) (Z^1+Z^2)+({\Re}(\mathcal{E})+1)(Z^1-Z^2)\right]\partial_\chi\nonumber\\
&+\frac{1}{2}\left[1-{\rm Im}(\mathcal{E})-{\Re}(\mathcal{E})+ Z^1 (Z^1+Z^2)+Z^2 (Z^1-Z^2)\right]\partial_{Z^1}\nonumber\\
&+\frac{1}{2}\left[1+{\rm Im}(\mathcal{E})-{\Re}(\mathcal{E})+Z^1 (Z^2-Z^1)+Z^2 (Z^1+Z^2)\right]\partial_{Z^2}\,,\nonumber\\
K_{J^2}&=\frac{(Z^2-Z^1)}{2}\partial_U+\frac{1}{2}\,\left({\rm Im}(\mathcal{E}) (Z^2-Z^1)+({\Re}(\mathcal{E})+1)(Z^1+Z^2)\right)\partial_\chi\nonumber\\
&+\frac{1}{2}\left(-{\rm Im}(\mathcal{E})+{\Re}(\mathcal{E})-1+ Z^1 (Z^2-Z^1)+Z^2 (Z^1+Z^2)\right)\partial_{Z^1}\nonumber\\
&+\frac{1}{2}\left(-{\rm Im}(\mathcal{E})-{\Re}(\mathcal{E})+1-Z^1 (Z^1+Z^2)-Z^2 (Z^1-Z^2)\right)\partial_{Z^2}\,,\nonumber\\
K_{J^3}&=-\frac{\chi}{2}\partial_U+\frac{1}{2}\,\left(-{\rm Im}(\mathcal{E})^2+{\Re}(\mathcal{E})^2-1\right)\partial_\chi\nonumber\\
&+\frac{1}{2}\left(-{\rm Im}(\mathcal{E})Z^1+Z^2({\Re}(\mathcal{E}))-3 Z^2\right)\partial_{Z^1}\nonumber\\
&+
\frac{1}{2}\left(-{\rm Im}(\mathcal{E})Z^2-Z^1({\Re}(\mathcal{E}))+3 Z^1\right)\partial_{Z^2}\,,\nonumber\\
K_{J_0}&=-\chi\partial_U+\left(-{\rm Im}(\mathcal{E})^2+{\Re}(\mathcal{E})^2-1\right)\partial_\chi+\nonumber\\
&+\left(-{\rm Im}(\mathcal{E})Z^1+Z^2({\Re}(\mathcal{E}))+ Z^2\right)\partial_{Z^1}+
\left(-{\rm Im}(\mathcal{E})Z^2-Z^1({\Re}(\mathcal{E}))- Z^1\right)\partial_{Z^2}\,,\nonumber\\
K_{H_0}&=\frac{1}{2}\partial_U+\chi\, \partial_\chi+\frac{1}{2}\,Z^i\,\partial_{Z^i}\,,\nonumber\\
K_{T_i}&=-\frac{1}{\sqrt{2}}\epsilon_{ij}\,Z^j\, \partial_\chi+\frac{1}{\sqrt{2}}\,\partial_{Z^i}\,,\nonumber\\
K_{T_\bullet}&=-\partial_\chi\,,
\end{align}
where we have used the following complex quantities:
\begin{equation}
\mathcal{E}=e^{2U}+|\mathcal{Z}|^2+i\,\chi\,\,;\,\,\,\mathcal{Z}=\frac{Z^1+i\,Z^2}{\sqrt{2}}\,.
\end{equation}
We give below the momentum maps restricted to the Cartan fields:
\begin{align}
\mathcal{P}_{J^1}^x&=(- 2 \cosh(U),0,0)\,\,;\,\,\,\mathcal{P}_{J^2}^x=(0,- 2 \cosh(U),0)\,\,;\,\,\,\mathcal{P}_{J^3}^x=(0,0,
\frac{1}{2}(3+\cosh(2U)))\,,\nonumber\\
\mathcal{P}_{J_0}^x&=(0,0,-2 \sinh^2(U))\,\,;\,\,\,\mathcal{P}_{H_0}^x=(0,0,0)\,\,;\,\,\,\mathcal{P}_{T_1}^x=\frac{\rme^{-U}}{\sqrt{2}}(-1,1,
0)\,\,;\,\,\,\mathcal{P}_{T_2}^x=-\frac{\rme^{-U}}{\sqrt{2}}(1,1,
0)\,,\nonumber\\
\mathcal{P}_{T_\bullet}^x&=(0,0,-\frac{1}{2}\,\rme^{-2U})\,.
\end{align}
\paragraph{The universal model manifold inside ${\rm G_{2(2)}/[SU(2)\times SU(2)]}$. } It is useful to define the embedding
\begin{equation}
\frac{{\rm SU}(2,1)}{{\rm U(2)}}\,\subset\,\,\frac{{\rm G}_{2(2)}}{{\rm SU(2)\times SU(2)}}\,.
\end{equation}
The scalars of the latter $\{U',\chi',y,\varphi,\zeta^\Lambda,\,\hat{\zeta}_\Lambda\}$ are expressed in terms of those spanning the former $\{U,\chi,Z^1,\,Z^2\}$ as follows:
\begin{align}
U'&=U\,\,,\,\,\,\chi'=\chi\,\,,\,\,\,y=\varphi=0\,,\nonumber\\
\zeta^0&=\frac{{Z^1}-{Z^2}}{2 \sqrt{2}}\,\,;\,\,\,\zeta^1=\frac{{Z^1}+{Z^2}}{2 \sqrt{2}}\,\,;\,\,\,
\hat{\zeta}_0=\frac{{Z^1}+{Z^2}}{2 \sqrt{2}}\,\,;\,\,\,\hat{\zeta}_1=-\frac{3 ({Z^1}-{Z^2})}{2 \sqrt{2}}\,.
\end{align}

\section{The \texorpdfstring{${\rm Sp(2,2)}/[{\rm USp(2)\times USp(2)}]$}{Sp(2,2)/[USp(2) x USp(2)]} model}\label{so14}
We use the description of the isometry group as ${\rm SO}(1,4)$. Its generators are written in the form:
\begin{equation}
\mathfrak{so}(1,4)={\rm Span}(J^x,\,J^{\prime x},\,H_0,\,N_i)\,,
\end{equation}
where $J^x$ are, as usual, the quaternionic structure generators. The matrix representation of the above generators is:
\begin{align}
J^x&=\left(\begin{matrix}0 & {\bf 0}\cr {\bf 0} & J_+^x\end{matrix}\right)\,\,;\,\,\,J^{\prime x}=\left(\begin{matrix}0 & {\bf 0}\cr {\bf 0} & J_-^x\end{matrix}\right)\,,\nonumber\\
H_0&=\left(
\begin{array}{lllll}
 0 & 1 & 0 & 0 & 0 \\
 1 & 0 & 0 & 0 & 0 \\
 0 & 0 & 0 & 0 & 0 \\
 0 & 0 & 0 & 0 & 0 \\
 0 & 0 & 0 & 0 & 0
\end{array}
\right)\,\,;\,\,\,Z^i\,N_i=\left(
\begin{array}{lllll}
 0 & 0 & Z^1 & Z^2 & Z^3 \\
 0 & 0 & Z^1 & Z^2 & Z^3 \\
 Z^1 & -Z^1 & 0 & 0 & 0 \\
 Z^2 & -Z^2 & 0 & 0 & 0 \\
 Z^3 & -Z^3 & 0 & 0 & 0
\end{array}
\right)\,,
\end{align}
where $J_\pm^x$ are the self-dual and anti-self-dual $4\times 4$ 't Hooft matrices.
The Borel subalgebra is generated by $\{H_0,\,N_i\}$ which satisfy the relations $[H_0,\,N_i]=N_i$, all other commutators being zero.
The parametrization id defined by the coset representative:
\begin{equation}
\mathbb{L}(q^u)=\rme^{Z^i\,N_i}\,\rme^{U\, H_0}\,.
\end{equation}
The metric in the normalization $\Omega^x_{uv}=-K^x_{uv}$ reads:
\begin{equation}
ds^2=\frac{1}{2}\left(dU^2+\rme^{-2U} (dZ^i)^2\right)\,.
\end{equation}
The scalar curvature being $\mathcal{R}=-24$.
\bibliography{supergravity}
\bibliographystyle{toine}
\end{document}